\documentclass[11pt]{article}
\usepackage{graphicx,amsmath,amsfonts,amssymb}
\usepackage{color,multicol,multirow}
\usepackage{array}
\def\sloppy{\tolerance=100000\hfuzz=\maxdimen\vfuzz=\maxdimen}
\def \beq  {\begin{equation}}
\def \eeq  {\end{equation}}
\def \beqar {\begin{eqnarray}}
\def \eeqar {\end{eqnarray}}
\def\bsp{\beq\begin{split}}
\allowdisplaybreaks
\mathchardef\mhyphen="2D

\def\vx {{\vec x}}

\def\vk {{\vec k}}
\def\vf {{\varphi}}

\def\Tr {{\rm Tr}}

\def\vk {\vec{k}}
\def\vx {{\vec x}}

\def\del {\partial}

\def\vf {{\varphi}}

\def\half{\textstyle{1\over 2}}

\begin{document}
\def \CMP {{Commun. Math. Phys.}}
\def \PRL {{Phys. Rev. Lett.}}
\def \PL {{Phys. Lett.}}
\def \NPBProc {{Nucl. Phys. B (Proc. Suppl.)}}
\def \NP {{Nucl. Phys.}}
\def \RMP {{Rev. Mod. Phys.}}
\def \JGP {{J. Geom. Phys.}}
\def \CQG {{Class. Quant. Grav.}}
\def \MPL {{Mod. Phys. Lett.}}
\def \IJMP {{ Int. J. Mod. Phys.}}
\def \JHEP {{JHEP}}
\def \PR {{Phys. Rev.}}
\begin{titlepage}
\null\vspace{-62pt} \pagestyle{empty}
\begin{center}
\rightline{CCNY-HEP-13/1}
\rightline{March 2013}
\vspace{1truein} {\Large\bfseries
Diffractive Effects and General Boundary Conditions}\\
\vskip .1in
{\Large\bfseries in Casimir Energy}\\
\vskip .1in
{\Large\bfseries ~}\\
{\large\sc Dimitra Karabali$^a$} and
 {\large\sc V.P. Nair$^b$}\\
\vskip .2in
{\itshape $^a$Department of Physics and Astronomy\\
Lehman College of the CUNY\\
Bronx, NY 10468}\\
\vskip .1in
{\itshape $^b$Physics Department\\
City College of the CUNY\\
New York, NY 10031}\\
\vskip .1in
\begin{tabular}{r l}
E-mail:&{\fontfamily{cmtt}\fontsize{11pt}{15pt}\selectfont dimitra.karabali@lehman.cuny.edu}\\
&{\fontfamily{cmtt}\fontsize{11pt}{15pt}\selectfont vpn@sci.ccny.cuny.edu}
\end{tabular}

\fontfamily{cmr}\fontsize{11pt}{15pt}\selectfont
\vspace{.8in}
\centerline{\large\bf Abstract}
\end{center}
The effect of edges and apertures on the Casimir energy of an arrangement of
plates and boundaries can be calculated 
in terms of an effective nonlocal
lower-dimensional field theory that lives on the boundary.
This formalism has been developed in a number of previous papers
and applied to specific examples with Dirichlet boundary conditions.
Here we generalize the formalism to arbitrary boundary conditions.
As a specific example, the geometry of a flat plate and a half-plate placed parallel to it is considered for a number of different boundary conditions and the area-dependent and edge dependent contributions to the Casimir energy are evaluated.
While our results agree with known results for those special cases (such as the Dirichlet and Neumann limits) for which other methods of calculation have been used, our formalism is suitable for general boundary conditions, especially for the diffractive effects.

\end{titlepage}

\pagestyle{plain} \setcounter{page}{2}
\setcounter{footnote}{0}
\setcounter{figure}{0}
\renewcommand\thefootnote{\mbox{\arabic{footnote}}}

\section{Introduction}
The Casimir effect \cite{Casimir}, the classic example of the influence of boundary conditions in quantum field theory, has been of considerable interest over the last several years. A vast body of literature has emerged with a number of new geometries being explored, both analytically and numerically and both at zero and nonzero temperatures \cite{reviews}.
Recent developments in nano-machinery have also provided further impetus to these efforts.
Geometries with edges and apertures are a particularly interesting set because of the possible interplay of diffractive effects and boundary conditions on the fields. The analytic calculation of the propagators in the given geometry with the given boundary conditions is rarely feasible and numerical and analytical approximations have to be used; diffractive effects are then a formidable computational task. We have recently developed a formalism which focuses on the diffractive effects
\cite{Kabat:2010nm}-\cite{proceedings}. In our approach, an effective nonlocal lower dimensional field theory defined on the boundaries, with a boundary action $S_B$, is first extracted. In the subsequent analysis of this lower dimensional theory, apertures and edges can be easily incorporated leading to a systematic calculational scheme for diffractive effects.
The results, both for zero and nonzero temperatures, were in good agreement with alternate methods of calculation, such as  the numerical world-line method \cite{{Gies}, {KlingGies}}, and the
scattering matrix method \cite{{MIT1},{MIT2}} and in those geometries where such calculations had been done. We also obtained a universal low energy theorem for the Casimir force between holes on a plate \cite{Kabat:2010yy}.

In this paper, we extend this formalism addressing two important and related issues. The boundary action $S_B$ is a functional of the field on the boundary and hence it is naturally defined for Dirichlet boundary conditions for the propagator or the fields. So the first question is: How do we apply our formalism to the case of Neumann or even more general boundary conditions? Although the case of two full parallel plates with no edge or diffractive effects for a scalar field with Robin boundary conditions has been studied before in \cite{Robin, cavalcanti}.  Our focus here is on the edge and diffractive effects.
While Dirichlet, Neumann or Robin are the easiest to work with, the general boundary conditions, say on a Laplace-type kinetic energy operator, are given by the von Neumann theory of self-adjoint extensions characterized by a self-adjoint operator on the space of boundary values
\cite{vonNeumann, asorey}.
However, there is a problem with the simple and straightforward generalization  to arbitrary boundary conditions because, generically, 
the (minus) Laplace operator has negative eigenvalues, suggesting an instability for the quantum theory \cite{asorey}.  On the other hand, physically, the boundary of interest is produced by the insertion of a material plate or some such object, and it would be rather strange if this process leads to a global instability. So the next question would be: How do we square these concerns? Further, it is not clear that, in a full fledged interacting quantum field theory, mimicking the material plate by boundary conditions is an adequate characterization. These are some of the issues discussed in this paper.

We start with a quick resume of the boundary action method. We then generalize the method to arbitrary boundary conditions in section 3. In section 4, we apply this to the case of two parallel plates with Robin boundary conditions, one of them being a semi-infinite plate so that edge effects and diffraction at the edge become important. The analytical part of the calculation is explained in some detail, followed by the numerical evaluation of some of the integrals needed. A variety of values of the Robin parameter are considered.
The case of two full plates (i.e., with no edge effects) with Robin boundary conditions \cite{Robin, cavalcanti} and
the geometry of a plate and half-plate but restricted to Neumann-Neumann conditions \cite{MIT1, MIT2} have been analyzed by other methods previously. We give a comparison of our results with these in the discussion section. Some other interesting features of our 
calculation are also pointed out there. The paper concludes with an appendix where some of the mathematical calculations we have done are outlined.


\section{Resum\'e of boundary action method}

\begin{figure}[!b]
\begin{center}
\scalebox{.8}{\includegraphics{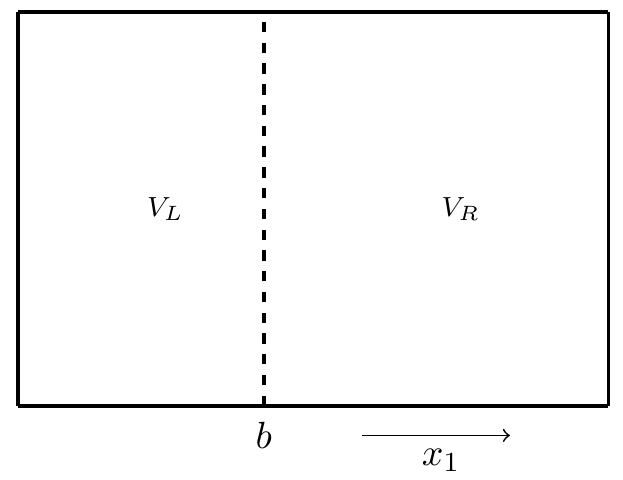}}
\caption{Schematics of splitting the functional integration into that over fields in two regions $V_L$ and $V_R$}
\label{pic1}
\end{center}
\end{figure}

We start with a brief summary of the approach developed in \cite{Kabat:2010nm, Kabat:2010yy, thermal}.
We will consider a scalar field theory with action $S(\phi)$ in a cubic box of volume $V$, with $V\rightarrow \infty$ eventually. For simplicity, we start off with a free field theory with a kinetic energy operator given by the Laplacian, i.e., $S(\phi ) = {\half} \int (\del \phi )^2$. The quantity of interest for the Casimir energy is the functional integral (or partition function)
\beq
Z = \int [d \phi ]\, ~\exp\bigl[ - S(\phi) \,\bigr]
\label{boun1}
\eeq
Our basic strategy is to consider the box as divided into a left region $V_L$ and a right region
$V_R$ as shown in Fig.\,\ref{pic1}, with an interface (shown as the dashed line) at, say,
$x_1 =b$. Then the functional integral can be done in two stages. We integrate over all field configurations in $V_L$ with a fixed value of the field (say, $ \phi =\vf$) on the interface and similarly for the right region $V_R$. Each integration leads to a functional of $\vf$, which
we will denote by $\Psi_L$ and ${\tilde \Psi}_R$ respectively, up to constant factors $Z_L$, $Z_R$, so that
\beqar
Z &=& Z_L\, Z_R\, \int [d\vf] ~\exp \bigl[ - S_B(\vf)\,\bigr]\nonumber\\
\Psi_L\, {\tilde \Psi}_R &=& \exp \bigl[- S_B(\vf)\,\bigr]
\label{boun2}
\eeqar
The final integration over the values of the field $\vf$ at the interface completes the functional integration in (\ref{boun1}). If there is a plate placed at $x_1 =b$ with the field vanishing on it, then this last integration is trivial; we just set $\vf =0$. If we have a plate with an opening, then $\vf =0$ everywhere on the interface except at the opening. The integration of $\exp(- S_B(\vf)\,)$ with $\vf$ restricted to being nonzero only at the opening gives a boundary contribution to the partition function, which will include all the diffractive effects due to the opening. This is our approach in a nutshell. It is clear that the formalism can be extended to boundaries with many components with different configurations of apertures, edges, etc. There will be many contributory terms to $S_B(\vf)$ and these would capture the various boundary effects.

This stage-wise functional integration is easily implemented by writing $\phi_L = \phi_{0L} + \eta_L$, where $\phi_{0L}$ is a specific field configuration in $V_L$ with the boundary value
$\phi_{0L} \rightarrow \vf$ as $x_1 \rightarrow b$. Explicitly, we may take it to be
\beq
\phi_{0L} (x) = \int_{\del V_L} \, \vf(x') \, ~n\cdot \del_{x'} \, G_L(x, x') 
\label{boun3}
\eeq
where $G_L(x, x')$ is the Green's function for $\square$ with Dirichlet boundary conditions, i.e.,
$\square G_L(x,x') = \delta^{(4)}(x-x')$, $G_L =0 $ if $x$ or $x' \in \del V_L$. In this equation,
$n\cdot \del$ denotes the derivative normal to the boundary $\del V_L$. The functional freedom for the fields in $V_L$ is in $\eta_L$ which is 
taken to vanish at all boundaries (including the apertures),
so that the full field $\phi_L$ has the required boundary behavior,
$\phi_L \rightarrow \vf$. This is consistent with the Dirichlet boundary condition for
the Green's function $G_L$, which is also the propagator for
$\eta_L$.
The functional integration over $\eta_L$ gives the partition function for the fields in $V_L$, namely, $Z_L$,
evaluated with the Dirichlet boundary condition, and also $\Psi_L$, which carries the
$\vf$-dependence. A similar splitting $\phi_R = \phi_{0R} + \eta_R$ can be used for the right region $V_R$. The boundary action is then found to be
\beq
S_B(\vf) = {1\over 2} \int \, \vf(x) ~ n\cdot \del_x \, ~n\cdot\del_{x'} \, \left[ G_L(x,x') + G_R (x,x')\right]~\vf(x')
\label{boun4}
\eeq 
where the integration is over the common interface and where $G_R$ also satisfies Dirichlet boundary conditions.

We see that the value of the field at the interface, namely, $\vf$ is what is left free until the last integration. It is easy to impose a vanishing condition on this field wherever there is material plate at the interface. However, imposing a Neumann condition (or anything more general) is 
not straightforward.
We may use a mode expansion of the fields in $V_L$ in terms of eigenmodes of $\square$,
obeying more general boundary conditions, but then extracting a boundary action 
(which can easily accommodate apertures) becomes awkward.
For example, if we want to use Neumann condition, we have to use bulk modes which obey this condition for the plate parts of the interface, but must have an arbitrary value $\vf$ on the openings, left free until the final integration. Finding such modes is a rather involved mathematical problem.
We would like a formalism which has the facility of dealing with apertures and diffractive effects as easily as we can do for the Dirichlet case, yet accommodate general boundary conditions directly in terms of the boundary action, so that the bulk modes are easy to construct.
In the next section, we will address this question.

Two remarks are useful at this point. Although we used a free scalar field theory to illustrate the set-up, it is clear that we can use a similar formalism for an interacting field theory as well. Secondly, if we think of the $x_1$-direction as time (with a Minkowski signature), the result of doing the functional integral with the specified value of the field, namely, $\vf$ at $x_1 =b$, would be a
wave functional (for some state) of the field theory. We have emphasized this by using the notation $\Psi_L (\vf)$. (If we take the left end of the box to be at $x_1 = - \infty$, then we get the ground state wave functional.) So what we have is really a Euclidean version of this wave functional set-up. As shown in \cite{Kabat:2010yy}, this wave functional can be obtained directly from the quantum effective action $\Gamma [\chi ]$. We solve $(\delta \Gamma_L /\delta \chi ) = 0$ subject to the boundary value $\chi \rightarrow \vf$
as $x_1 \rightarrow b$. Then $\Psi_L = \exp (- \Gamma^*_L)$ where
$\Gamma^*_L$ is given by $\Gamma [\chi ]$ evaluated on this solution.
(Obviously, similar statements hold for ${\tilde \Psi}_R$ as well.) This gives another way to think of our procedure in a fully interacting quantum field theory. 
\section{General boundary conditions}

We start again with the simple case of a free scalar field and consider the von Neumann theory of self-adjoint extensions \cite{vonNeumann}.
This has recently been rephrased nicely by Asorey, Ibort and Marmo, so we can use their approach \cite{asorey}.
For the Laplace operator in some region $V$, we consider the field $\vf$ and its normal derivative
$\del_n \vf$. On the boundary, we consider square integrable functions, i.e.,  they form a Hilbert space. Thus the combination $\vf +i \del_n \vf$ (where $\del_n = n\cdot \del$ is the normal derivative) may be viewed as an element of this
Hilbert space.
The most general boundary condition, according to the von Neumann theory, is
\beqar
\vf +i \del_n \vf &=& U \, ( \vf - i \del_n \vf )\nonumber\\
(\vf +i \del_n \vf) (x) &=& \oint_y U(x,y)  \, ( \vf - i \del_n \vf )(y)\label{1}
\eeqar
where $U$ is a unitary operator on the boundary Hilbert space, namely, on the space of
${\mathbb L}^2$-functions on the boundary, of which $\vf$ is an element.
In the second line of the equation above, we emphasize this by writing $U(x, y)$.
Using the operator notation of the first line of (\ref{1}),
we find
\beq
\del_n \vf = -i \left( {U -1 \over U+1}\right)\, \vf \equiv - {\cal K} \, \vf \label{2}
\eeq
${\cal K}$ is a hermitian operator and the transformation from $U$ to ${\cal K}$ is the so-called Cayley transform. There are two interesting limits, ${\cal K} \rightarrow 0$ and
${\cal K} \rightarrow \infty$. The first one, equivalent to $U =1$, gives the Neumann boundary condition and the second one ($U \rightarrow -1$) corresponds to the Dirichlet condition.
(For the latter, it is better to divide (\ref{2}) by ${\cal K}$ and then take the limit.)
These are special points. The case of ${\cal K}$ being a constant (proportional to the identity
on the Hilbert space) is the Robin condition.
The general theorem is that, in the space of ${\cal K}$'s, infinitesimally close to the two limits of Neumann and Dirichlet, 
the Laplace operator can have negative eigenvalues \cite{asorey}.

The obvious question is whether this can affect our evaluation of the functional integral.
To ensure consistency of the variational problem with the boundary conditions, extra surface terms may have to be added to the action, and one can ask if they are sufficient to avoid any pathologies. The answer, in general, is that the negative eigenvalues can lead to pathologies; 
but for certain types of ${\cal K}$, or range of eigenvalues for the same, we can have a stable situation. This is further commented upon in the discussion section.

There is a different way to look at this problem, which also suggests the solution. Let us go back again to (\ref{boun2}) writing the partition function as
\beqar
Z &=& [\det (-\square_L)]^{-\half} ~[\det (-\square_R)]^{-\half} ~\int [d\vf ]~
\Psi_L(\vf) \, {\tilde \Psi}_R (\vf)\nonumber\\
&\equiv& Z_L\, Z_R\, Z_B\label{17}
\eeqar
If the partition between $V_L$ and $V_R$ is real, like a plate, may be with openings, 
we express that as the vanishing of the field
$\phi = \vf$ in the plate region of the partition, but free on the open region.
Equivalently, we may say that the plate is represented by
$\delta (\vf)$ where the delta function is only on the plate region.  Explicitly, we can expand
$\vf$ in term of modes on the open regions of the interface, and on the plate parts. The coefficients of the latter vanish by the delta function. The result then agrees with what we did in \cite{Kabat:2010nm}.
We may interpret this delta function as the operator representing the insertion of the plate, the plate operator.
So the final integral, apart from the determinants,  looks like
\beq
Z_B = \int [d\vf ]~
\Psi_L(\vf) \, \Bigl[\delta_{plate} (\vf)\Bigr]\, {\tilde \Psi}_R(\vf)\label{18}
\eeq
Now consider what happens when we have a normal derivative.
The normal derivative acts as a functional derivative on the $\Psi$'s. This can be seen easily from slicing up the path integral along the direction normal to the interface, the $x_1$ direction. The Euclidean action then looks like
\beq
S = S( \{ \vf_i \}) = {1\over 2} \int d^3x^T~ \left[ {(\vf_N - \vf_{N-1})^2\over x_N - x_{N-1}}
+ {(\vf_{N-1} - \vf_{N-2})^2\over x_{N-1} - x_{N-2}} + \cdots + (\nabla^T\vf)^2\right]
\label{19}
\eeq
$\vf_N$ is the boundary value $\vf$. The boundary action is given by
\beq
\Psi (\vf ) \equiv e^{-S_B(\vf)} = \int \prod_1^{N-1} d\vf_i  \, ~\exp (- S (\{\vf_i \})
\label{19a}
\eeq
Differentiating $e^{-S}$ with respect to
$\vf_N$, we find
\beqar
{\delta \over \delta \vf_N} ~e^{-S} &=& -{(\vf_N - \vf_{N-1} ) \over x_N - x_{N-1}}~e^{-S}\nonumber\\
&\rightarrow& - \del_n \vf ~ e^{-S}
\label{20}
\eeqar
So, going back to (\ref{17}, \ref{18}), we see that imposing a boundary condition
$\vf' + {\cal K} \, \vf =0$ on the plate is equivalent to imposing
\beq
\left[ - {\delta \over \delta \vf} ~+ {\cal K} \, \vf \right] ~\Psi(\vf)  = 0
\label{21}
\eeq
To see how this requirement can be obtained, consider
the integral
\beq
I = \int [d\vf] \, \exp\left( {-{1\over 2} \int \vf\, {\cal K}\,  \vf }\right) \,\, \Psi[\vf]
\label{21a}
\eeq
We can then write
\beqar
 \int [d\vf] \, \exp\left( {-{1\over 2} \int \vf\, {\cal K}\,  \vf }\right) \,\, (\del_n \vf)\, \Psi[\vf]
&=&  \int [d\vf] \, \exp\left( {-{1\over 2} \int \vf\, {\cal K}\,  \vf }\right) \,\, \left( -{\delta \over \delta \vf}\right)\,\Psi[\vf]\nonumber\\
&=& \int [d\vf] \, \exp\left( {-{1\over 2} \int \vf\, {\cal K}\,  \vf }\right) \,(- {\mathcal K}\, \vf )\, \Psi[\vf]
\label{21b}
\eeqar
where, in the second line, we have done a partial integration. This equation can be rewritten as
\beq
 \int [d\vf] \, \exp\left( {-{1\over 2} \int \vf\, {\cal K}\,  \vf }\right) \,\, (\del_n \vf  + {\mathcal K}\vf)\, \Psi[\vf]
= 0 \label{21c}
 \eeq
 Thus we do obtain the required vanishing of $\del_n \vf + {\mathcal K}\vf $, showing that the operator representing the plate must be taken as
 $\exp (-{\half} \int \vf {\mathcal K}\vf)$. The boundary condition applies only for the fields on the plate part of the boundary, so we must restrict the fields in $ \int \vf {\mathcal K}\vf$ to be only on the plate part.
Further, we can choose the boundary conditions independently on the left and right sides of the same plate, which have independent boundary fields $\vf_L$ and $\vf_R$, so that the insertion of  $\exp (-{\half} \int \vf {\mathcal K}\vf)$ is to be done separately for each. (This issue did not arise in previous calculations because we used Dirichlet conditions setting
$\vf=0$ for both sides of the plate.) The fields on the aperture part are the same on both sides of the boundary.

So the calculational algorithm is: Use Dirichlet conditions with left and right regions and fields
$\phi_L$ and $\phi_R$. These go to the boundary values
$\vf_L$ and $\vf_R$. Then calculate
\beqar
Z_B &=& \int [d\vf_L \, d \vf_R] ~\delta_{aperture}(\vf_L - \vf_R)~
\exp\left( - {1\over 2}\int_{plate,L} \vf_L {\mathcal K}_L\,\vf_L  - {1\over 2}\int_{plate,R} \vf_R {\mathcal K}_R\,\vf_R \right) \nonumber\\
&&\hskip .5in \times  \Psi[\vf_L] \, \Psi[\vf_R]
\label{23}
\eeqar
We integrate over all fields $\vf_L$ and $\vf_R$, but the $\delta$-function ensures that
the fields are the same on both sides of the aperture and the integrals
in the exponents are restricted to the fields on the plate part of the boundary on the left and right sides.
This formulation, at least partially, avoids the problem of potential negative eigenvalues since the bulk determinants leading to
$Z_L$ and $Z_R$ are always calculated with Dirichlet conditions.

For the free theory, $\Psi (\vf )$ is of the form
\beq
\Psi (\vf ) = \exp \left( - {\half} \int \vf (x) \, M(x,x') \, \vf (x') \right)
\label{26}
\eeq
where $M(x,x') = n\cdot \del_x \, n\cdot \del_{x'} \, G(x, x')$. Thus, in this case,
 the expression
(\ref{23}) for $Z_B$ becomes
\beqar
Z_B &=& \int [d\vf_L \, d \vf_R] ~\delta_{aperture}(\vf_L - \vf_R)\,
e^{- S_B}\nonumber\\
S_B&=&  {1\over 2} \int_{plate, L} \vf_L\, {\mathcal K}_L \,\vf_L 
+ {1\over 2} \int_{plate, R} \vf_R\, {\mathcal K}_R \,\vf_R 
+ {1\over 2} \int_{boundary} \vf \, M \, \vf
\label{27}
\eeqar
Once again, the integrals involving ${\mathcal K}$'s are only over the plate regions of the boundary, while the last integral is over all boundary.
If the boundary has disconnected components, as would be the case for, say, parallel plates,
then all such components must be included in (\ref{27}). We are now in a position to apply this method of calculation to a specific example.

\section{Plate and Half-plate}
\subsection{Modes on the boundary and the boundary action}

\begin{figure}[!b]
\begin{center}
\scalebox{.9}{\includegraphics{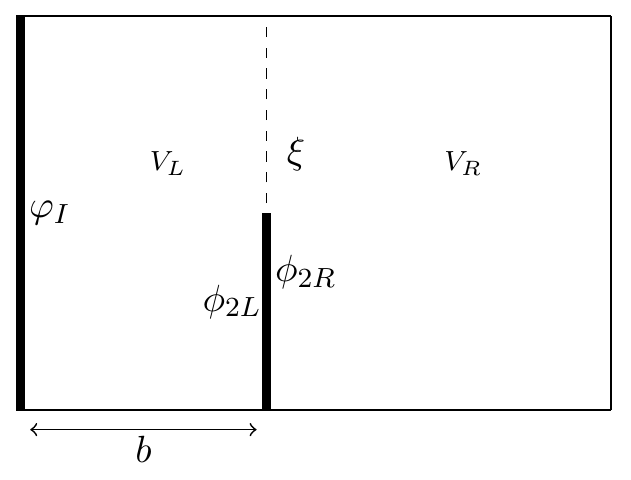}}
\caption{The arrangement of a Robin plate and a Robin half-plate with values of fields indicated}
\label{pic3}
\end{center}
\end{figure}

We will consider the arrangement shown in Fig.\,\ref{pic3} where we have an
infinite plate and a semi-infinite plate parallel to it, separated in the $x_1$-direction by a distance $b$. 
The case when the field has Dirichlet boundary conditions with $\phi = 0$ on the plates was analyzed previously \cite{Kabat:2010nm}. Here we will consider more general boundary conditions. The bulk contribution from the left and right regions, namely, $Z_L\, Z_R$, which in the present case is also calculated with Dirichlet boundary conditions, is the same as before. For the boundary contributions, first of all, we need the boundary action. The Dirichlet propagator in the left region, between the two plates, is given by
\beqar
G(x, x') &=& \int {d^3 k \over (2\pi )^3} \, e^{i k \cdot (x^T- x'^{T})}\, G_D (\omega , x_1 , x_1')\nonumber\\
G_D (\omega, x_1, x_1') &=& {1\over 2 \omega}\, N(\omega)\, \left[ \theta (x_1 - x_1') \, \left( e^{\omega( x_1 - x_1')}
+  e^{2 b \omega}\, e^{- \omega( x_1 - x_1')}\right)\right.\nonumber\\
&&\hskip .7in  \theta (x_1' - x_1) \,  \left( e^{-\omega( x_1 - x_1')}
+  e^{2 b \omega}\, e^{ \omega( x_1 - x_1')}\right)\nonumber\\
&&\hskip .7in \left. - \,\left( e^{\omega (x_1 + x_1')} + e^{2 b \omega} \, e^{-\omega (x_1 +x_1')}\right)
\right] \label{29}
\eeqar
where $\theta (x_1 - x_1')$ is the step function and 
\beq
N(\omega ) = {1\over e^{2 b \omega} - 1} \label{30}
\eeq
Further, $\omega = \sqrt{k^2 } = \sqrt{k_0^2 +k_2^2 +k_3^2}$,~ $x^T = ( x_0, x_2, x_3)$. We will take the open part
of the second plate to be in the range
$x_2 \geq 0$.
Calculating the normal derivatives, we find
\beqar
S_B &=& {1\over 2} \int \vf_I(x) \left( \omega \coth b \omega + {\cal K}_I \right)_{x,x'}\,\vf_I (x')
+ \int \vf_I (x) \left( - {\omega\, \text{\,csch\,} b  \omega} \right)_{x, x'}\, \vf_{II}(x')\nonumber\\
&&\hskip .1in + {1\over 2} \int \vf_{II}(x) \,\left( \omega \coth b \omega + {\cal K}_{L} \right)_{x,x'}\,\vf_{II}(x') + {1\over 2} \int \vf_R (x) \, \left( \omega + {\cal K}_R \right)_{x, x'} \vf_R(x')\nonumber\\
\label{31}
\eeqar
where $\vf_I$ refers to the field on plate I, at $x_1 =0$, and $\vf_{II}$ to that on the left side
($V_L$ side)
of the interface at $x_1 = b$ (which includes the half-plate and aperture) and $\vf_R$ to the field again at the same interface, but on the right side (the $V_R$ side).

We will now need to specify the operators ${\mathcal K}_I$, ${\mathcal K}_{L}$ and
${\mathcal K}_R$. For illustrating the techniques outlined, we will choose a simple Robin boundary condition ${\mathcal K}_I = \kappa_I =$ constant for the left plate (plate I).
The special cases $\kappa_I =0$ and $\kappa_I \rightarrow \infty$ will correspond to
the Neumann and Dirichlet limits for the left plate.
To specify ${\mathcal K}_{L}$ and ${\mathcal K}_R$, it is useful to separate $\vf_{II}$ into modes which have support on the aperture, modes with support on the plate, and the value of the field
at the edge of plate II, $x_2 = 0$. This is done by writing
\beqar
\vf_{II}&=& \int {d^3 k \over (2\pi)^3} e^{ik\cdot x} \left[ 
\int_0^\infty {dp \over \pi} {2\, p \over p^2 - (k_2 - i \epsilon)^2} \, \xi(\vk, p)
- \int_0^\infty {dp \over \pi} {2\, p \over p^2 - (k_2 + i \epsilon)^2} \, \phi_{2L}(\vk, p)\right.
 \nonumber\\
&&\hskip 1.2in \left.
+ { 2 \, \vert\vk\vert \over k_2^2 + \vk^2} \, \rho (\vk )
\right]\label{PHP1}
\eeqar
Here $\vk = (k_0, k_3)$, the two directions transverse to the edge of the plate, as well as
transverse to the plate itself. We define the fields
\beqar
\phi_{2L} (x) &=& \int {d^2 k \over (2\pi )^2} \int_0^\infty 2 {dp \over \pi} \, e^{i \vk \cdot \vx} \sin(p x_2)~\phi_{2L}(\vk, p)\nonumber\\
\xi (x) &=& \int {d^2 k \over (2\pi )^2} \int_0^\infty 2 {dp \over \pi} \, e^{i \vk \cdot \vx} \sin(p x_2)~\xi(\vk, p)\nonumber\\
\rho(x) &=& \int  {d^2 k \over (2\pi )^2} e^{i \vk \cdot \vx}\, \rho (\vk )\,
\left\{ \begin{matrix}
e^{-\vert\vk\vert x_2} & ~~~& x_2 > 0\\
e^{\vert\vk\vert x_2} & ~~~& x_2 < 0\\
\end{matrix}\right.
\label{PHP2}
\eeqar
The field $\xi$ is the field at the aperture ($x_2 > 0$) but vanishing at the edge $x_2 =0$,
$\phi_{2L}$ is the field on the plate ($x_2 <0$), vanishing at $x_2 =0$ and $\rho$ is essentially the value of the field at $x_2 =0$, but continued in a very specific way, with no additional functional degrees of freedom, to $x_2 >0$ and $x_2 <0$. (While the
value of the field at $x_2 = \pm \infty$ will not be relevant for us, the value of the field vanishing
at $x_2 = \pm \infty$ will make some of the integrals easier and better defined.
The continuation of the field value at the edge, namely $\rho$, to other values of
$x_2$  has been done in one particular way which ensures this.)

Carrying out the $p$-integration in (\ref{PHP1}) shows that
\beq
\vf_{II}(x) = \left\{ \begin{matrix}
\xi (x) + \rho (x)~~~~~ & ~~~&x_2 >0\\
\phi_{2L}(x) + \rho(x) ~~&~~~&x_2<0\\
\rho(x_0, x_3, x_2=0) &~~~&x_2 =0\\
\end{matrix}\right.
\label{PHP3}
\eeq
The amplitudes of the modes in (\ref{PHP1}), namely, $\xi (\vk, p)$, $\rho(\vk)$, and
$\phi_{2L}(\vk, p)$ constitute the functional freedom in the value of the field at the 
boundary.
In a similar way, we can write
\beq
\vf_{R}(x) = \left\{ \begin{matrix}
\xi (x) + \rho (x)~~~~~ & ~~~&x_2 >0\\
\phi_{2R}(x) + \rho(x) ~~&~~~&x_2<0\\
\rho(x_0, x_3, x_2=0) &~~~&x_2 =0\\
\end{matrix}\right.
\label{PHP3a}
\eeq
where
\beq
\phi_{2R} = \int {d^2 k \over (2\pi )^2} \int_0^\infty 2 {dp \over \pi} \, e^{i \vk \cdot \vx} \sin(p x_2)~\phi_{2R}(k)
\label{PHP3b}
\eeq
For $\vf_I$, since there is a single plate for all $x_2$, we can use a simple
mode expansion,
\beq
\vf_I = \int {d^3k \over (2\pi )^3} \, e^{ik\cdot x} ~\phi_1 (k)
\label{PHP4}\\
\eeq
The normalization for $\sin(p x_2)$ is
\beq
\int_0^\infty dx_2~ \sin(p x_2) \, \sin (q  x_2) = {\pi \over 2} \delta (p - q)
\label{PHP5}
\eeq
One advantage in parametrizing the fields as in (\ref{PHP1}) is that this already takes care of
the $\delta$-function $\delta_{aperture}(\vf_{II} - \vf_R)$ enforcing equality of fields
on the aperture $x_2 \geq 0$.

We may now regard ${\mathcal K}_{L}$ as an operator on the fields $(\phi_{2L}, \rho )$ and likewise ${\mathcal K}_R$ as an operator on $(\phi_{2R} , \rho )$.
While $\xi$ and $\phi_{2L}$ (and $\xi$ and $\phi_{2R}$) are orthogonal to each other,
$\rho$ is not orthogonal to $\xi$ or $\phi_{2L}$ (or $\phi_{2R}$).
Nevertheless, $(\xi, \phi_{2L}, \rho )$ and
$(\xi, \phi_{2R}, \rho )$ form a complete basis for fields at
$x_1 = b$. There is clearly an infinity of choices possible for ${\mathcal K}_L$ and
${\mathcal K}_R$, but, once again, for our analysis, we will make a simple choice.
We will take ${\mathcal K}_L = \kappa_L =$ constant for all the modes
$\phi_{2L}$ and ${\mathcal K}_R = \kappa_R =$ constant for all the modes
$\phi_{2R}$. In other words, the operators
${\mathcal K}_L$ and ${\mathcal K}_R$ are diagonal with the same eigenvalue $\kappa_L$
(respectively $\kappa_R$) for all $\phi_{2L}$ (respectively $\phi_{2R}$).
This is almost like a Robin condition for the plate on the right as well.
We say ``almost" because the situation with $\rho$ is a bit tricky.
It corresponds to the field just at the edge $x_2 =0$.
In principle, the value of ${\mathcal K}_L$ for $\rho$  can be different from the values of the same operator for $\phi_{2L}$, even after we have 
chosen the latter to be the same for all modes $\phi_{2L}$. A similar statement applies to
${\mathcal K}_R$ for $\rho$ versus ${\mathcal K}_R$ for $\phi_{2R}$.
In addition, we can have nondiagonal terms mixing $\rho$ with $\phi_{2L}$ and
$\phi_{2R}$. Thus even with the simplifying choices we have made, there are many parameters specifying the boundary behaviour. Summarizing, the boundary action (\ref{27}) for this geometry, with the choices we have made,  is
\beqar
S_B &=& {1\over 2} \int \vf_I (x) \, (\omega \coth b \omega \, +\kappa_I )_{x,x'} \vf_I(x')
+ \int \vf_I (x) (- \omega {\rm csch} b\omega )_{x,x'} \vf_{II} (x')\nonumber\\
&&+{1\over 2} \int \vf_{II} (x) \,(\omega \coth b\omega )_{x,x'} \vf_{II}(x') + {1\over 2}
\int \vf_R(x)\, (\omega )_{x,x'} \vf_R(x')\nonumber\\
&&+{1\over 2} \kappa_L \int \phi_{2L}\, \phi_{2L} + 
{1\over 2} \kappa_R \int \phi_{2R}\, \phi_{2R}  + {1\over 2} c_0  \int \rho \, \rho\nonumber\\
&&+ c_{2L} \int \phi_{2L}\,\rho
+  c_{2R} \int \phi_{2R}\, \rho 
\label{PHP5a}
\eeqar

\subsection{Simplification of the boundary action}

We are now in a position to evaluate the action in terms of the mode expansions we have given
and then carry out the functional integrals. We will only give
 the final form of the boundary action here, relegating the details to the appendix.
 
We will separate the terms into two categories. Because $\rho(\vk)$ depends only on two dimensions, it is easier to integrate out the other fields first and leave the $\rho$ integral to the end. Towards this, we shall first simplify the terms involving $\vf_I$, $\phi_{2L}$, $\phi_{2R}$ and $\xi$.
In this way, we separate out the terms in (\ref{PHP5a}) as
\beqar
S_B &=& S_B^{(1)}~+~ S_B^{(2)}\nonumber\\
S_B^{(1)}&=& {1\over 2}  \int \phi_1 \,{\mathcal M}_1\, \phi_1 
+ \int \phi_1 N_{1\,2L}\, \phi_{2L} + \int \phi_1 \, N_{1\,\xi} \, \xi + {1\over 2} \int \xi \, {\mathcal M}_{\xi} \, \xi \nonumber\\
&&\,
+{1\over 2} \int \phi_{2L} \, {\mathcal M}_{2L} \, \phi_{2L} 
+ {1\over 2} \int \phi_{2R} \, {\mathcal M}_{2R} \, \phi_{2R}
+\int \phi_{2L}\, Q_{2L \xi} \, \xi + \int \phi_{2R} \, N_{2R \xi} \, \xi
\label{PHP5b}\\
S_B^{(2)}&=& {1\over 2} \int \rho\, {\mathcal M}_\rho \, \rho + \int \phi_1 \, N_{1\rho} \, \rho
+ \int \phi_{2L} \, Q_{2L \rho} \, \rho + \int \phi_{2R} \, Q_{2R  \rho} \, \rho
+\int \rho\, Q_{\rho\xi} \, \xi
\label{PHP5c}
\eeqar
The various coefficient functions, such as ${\cal M}_1$, $N_{1\,2L}$, etc., arise naturally from the restriction of the coefficient functions in (\ref{PHP5a}) to the appropriate modes. We do not give their expressions here; they are given in the appendix.
Integrating over $\phi_1$, $\phi_{2L}$ and $\phi_{2R}$ yields
\beqar
-\log Z_B&=& {1\over 2} \Tr\,\log (\omega \coth b\omega \, +\kappa_I ) + 
{1\over 2} \Tr\, \log ( {\mathbb M}_L) + {1\over 2} \Tr\, \log ({\mathbb M}_R)
-\log Z'_B\label{PHP5d1}\\
Z'_B &=& \int [d\xi d\rho] \, e^{-S'_B}\nonumber\\
S'_B&=& {1\over 2} \int_{\vk,p,q}
~\xi(-\vk, p) \,{\mathbb M}_\xi (p,q) \,\xi(\vk, q)\nonumber\\
&&\hskip .2in +\int_{\vk,p} \rho (-\vk) \, N_{\rho\xi}(\vk,p)
\, \xi (\vk,p) + {1\over 2} \int_{\vk}
~ \rho (-\vk) \,{\mathbb M}_\rho\, \rho (\vk)\nonumber\\
&& \hskip .2in - {1\over 2}\int_{\vk,p,q} \left\{ \left[ \int_{p'} \,N_{2L\xi}(\vk,p,p') \,\xi(-\vk, p') - 
\rho (-\vk) \, N_{2L\rho} (\vk,p)\right] \,G_{2L}(\vk,p,q) \right. \nonumber\\
&&\hskip 1in\left. \times \left[ \int_{q'} \,N_{2L\xi}(\vk,q,q') \,\xi(\vk, q') - 
\rho (\vk) \, N_{2L\rho} (\vk,q)\right]\right\} \nonumber\\
&& \hskip .2in - {1\over 2}\int_{\vk,p,q} \left\{ \left[ \int_{p'} \,N_{2R\xi}(\vk,p,p') \,\xi(-\vk, p') + 
\rho (-\vk) \, N_{2R\rho} (\vk,p)\right] \,G_{2R}(\vk,p,q) \right. \nonumber\\
&&\hskip 1in\left. \times \left[ \int_{q'} \,N_{2R\xi}(\vk,q,q') \,\xi(\vk, q') + 
\rho (\vk) \, N_{2R\rho} (\vk,q)\right]\right\} 
\label{PHP5d2}
\eeqar
The indicated integrations are done with the measures
\beq
\int_{\vk, p,q} = \int {d^2k \over (2\pi)^2} {dp\over \pi} {dq\over \pi} ,\hskip .3in
\int_{p'} = \int_0^\infty {dp' \over \pi}, \hskip .2in {\rm etc.}
\label{PHP5e}
\eeq
The other quantities in (\ref{PHP5d1}, \ref{PHP5d2}) are
\beqar
{\mathbb M}_L &=&  2\pi\, (H_p+\kappa_L) \, \delta (p-q) + (\Delta {\mathbb M}_L)_{pq}
\label{PHP5f}\\
(\Delta {\mathbb M}_L)_{pq}&=& 4 pq\,\int_0^\infty {ds\over \pi} 
\left[ {H_s \over (p^2 - s^2) (q^2 - s^2)} + {H_p \over (s^2 - p^2) (q^2 - p^2)}  +{H_q \over (s^2 - q^2) (p^2 - q^2)}  \right]
\nonumber
\eeqar
\beqar
{\mathbb M}_R &=&  2\pi\, (\omega_p +\kappa_R)  \, \delta (p-q) + (\Delta {\mathbb M}_R)_{pq}
\label{PHP5g}\\
(\Delta {\mathbb M}_R)_{pq}&=& 4 pq\,\int_0^\infty {ds\over \pi} 
\left[ {\omega_s \over (p^2 - s^2) (q^2 - s^2)} + {\omega_p \over (s^2 - p^2) (q^2 - p^2)}  +{\omega_q \over (s^2 - q^2) (p^2 - q^2)}  \right]
\nonumber
\eeqar
\beqar
{\mathbb M}_\xi &=& 2\pi\, (H_p + \omega_p) \delta(p- q) + (\Delta {\mathbb M}_\xi)_{pq}
\label{PHP5h}\\
(\Delta {\mathbb M}_\xi)_{pq}&=& 4 pq\,\int_0^\infty {ds\over \pi} 
\left[ {(H_s+\omega_s) \over (p^2 - s^2) (q^2 - s^2)} + {(H_p+\omega_p) \over (s^2 - p^2) (q^2 - p^2)}  +{(H_q+\omega_q)\over (s^2 - q^2) (p^2 - q^2)}  \right]
\nonumber\\
&&~\nonumber\\
{\mathbb M}_\rho &=&  4 \vert\vk\vert^2 \,\int_{-\infty}^\infty {ds \over 2\pi} \,  {H_s+\omega_s \over \omega_s^4}
~+~ {c_0 \over 2 \vert \vk\vert} \label{PHP5i}
\eeqar
In these formulae,
\beq
H_p = \omega_p \left( { \omega_p + \kappa_I \coth b \omega_p \over \omega_p \coth b\omega_p + \kappa_I}\right), \hskip .2in \omega_p^2 = \vk^2 + p^2\label{PHP5j}
\eeq
The quantities $G_{2L}$ and $G_{2R}$ are the Green's functions or inverses of
the kernels ${\mathbb M}_L$ and ${\mathbb M}_R$ respectively. Explicitly, they are of the form
\beqar
G_{2L} (p,q) &=& \pi \, \delta(p - q) \,{1\over 2 (H_p + \kappa_L)}
- {1\over 2 (H_p + \kappa_L)} (\Delta{\mathbb M}_L)_{pq} \, {1\over 2 (H_q + \kappa_L)}\nonumber\\
&&+ \int_0^\infty {dp'\over \pi}  {1\over 2 (H_p + \kappa_L)} (\Delta{\mathbb M}_L)_{pp'} \, {1\over 2 (H_{p'} + \kappa_L)}
 (\Delta{\mathbb M}_L)_{p' q} \, {1\over 2 (H_q + \kappa_L)}
+\cdots\label{PHP5k}\\
G_{2R} (p,q) &=& \pi \, \delta(p- q) \,{1\over 2 (\omega_p + \kappa_R)}
- {1\over 2 (\omega_p + \kappa_R)} (\Delta{\mathbb M}_R)_{pq} \, {1\over 2 (\omega_q + \kappa_R)}\nonumber\\
&&+ \int_0^\infty {dp'\over \pi}  {1\over 2 (\omega_p + \kappa_R)} (\Delta{\mathbb M}_R)_{pp'} \, {1\over 2 (\omega_{p'} + \kappa_R)}
 (\Delta{\mathbb M}_R)_{p' q} \, {1\over 2 (\omega_q + \kappa_R)}
+\cdots \label{PHP5m}
\eeqar
Finally, we also have
\beqar
N_{\rho\xi} (\vk,p)&=& 4 \vert\vk\vert\, p \int_0^\infty {ds \over \pi} 
\left[ {(H_s +\omega_s) \over \omega_s^2} - {(H_p +\omega_p)  \over \omega_p^2} \right] {1\over p^2 - s^2}\nonumber\\
N_{2L\xi}(\vk, p,p') &=& -(\Delta {\mathbb M}_L)_{pq} \nonumber\\
N_{2R\xi} (\vk, p,p')  &=& -(\Delta {\mathbb M}_R)_{pq} 
={4\, p\, p'\over (p'^2- p^2)}
\left[ (p'^2 + \vk^2 )\, {\cal I}(\vk, p') - (q^2 + \vk^2 )\, {\cal I}(\vk, q) \right]
\label{PHP5n}\\
N_{2L\rho}(\vk, p)&=& 4 \vert\vk\vert \, p \left[ \int_0^\infty {ds \over \pi} \left(
{H_s \over \omega_s^2} - {H_p \over \omega_p^2} \right) {1\over p^2 - s^2}
~+~ {c_{2L}\over 2 \vert\vk\vert \, \omega_p^2}\right]
\nonumber\\
N_{2R\rho}(\vk, p)&=& - 4 \vert\vk\vert \, p \left[ {\cal I}(\vk,p) 
~+~ {c_{2R}\over 2 \vert\vk\vert \, \omega_p^2}\right]
\label{PHP5p}
\eeqar
where
\beq
{\cal I}(\vk ,p)= {1\over 2\pi\,p \sqrt{\vk^2 +p^2}} \log \left( {\sqrt{\vk^2 + p^2}\,+p \over
\sqrt{\vk^2 + p^2}\,-p }\right)\label{PHP5q}
\eeq
The equalities $N_{2L\xi}=-\Delta{\mathbb M}_L$ and $N_{2R\xi}=-\Delta{\mathbb M}_R$ show that $N_{2L\xi}$, $N_{2R\xi}$ are of the same
order in diffractive effect as the $\Delta M$'s, an issue which is relevant when we do the expansions explained below.

\subsection{Evaluation of the Casimir energy}

We are now in a position to write down the free energy,
which will have several terms corresponding to the contributions from the 
left and right bulk regions and 
integrations over $\phi_1$, $\phi_{2L}$, $\phi_{2R}$, etc. First, 
in extracting the free energy, we note that the $\vk$-dependence is the same for 
all terms and that the overall integration factor representing the trace is
$L_0 L_3 ({d^2 k / (2\pi )^2}) $. With $-\log Z_B = L_0\, F$, the free energy is given by
\beq
F= F_{bulk} ~+~ F_{I}  ~+~F_{2L} ~+~F_{2R}~+~ F_{\xi} ~+~ F_{\rho}\label{PHP5r}
\eeq
We have added in $F_{bulk}$, which is the contribution from the bulk determinants 
(from $Z_L, \, Z_R$), calculated with Dirichlet conditions.  The remaining terms in (\ref{PHP5r}) arise
from the boundary action and correspond to the terms on the right hand side of
(\ref{PHP5d1}) or the result of integrating over the terms
on the right hand side
of (\ref{PHP5d2}). All these contributions have divergences corresponding to
free space with no plates. They can be identified as the $b\rightarrow \infty$ limit
of the expressions here. Thus, in the following calculations, the renormalization of all terms will be done by subtracting
the $b\rightarrow \infty$ limit.

\subsubsection*{\underline{A note on the expansion scheme}}

The exact calculation of the relevant determinants and the free energy is still very involved and an expansion scheme is needed to get a good approximation for several of the terms
in (\ref{PHP5r}). In our earlier work \cite{Kabat:2010nm}, we noticed that the relevant operator,
say ${\mathbb M}$, had the structure ${\mathbb M}_{pq} = {\mathbb M}^{(0)} \delta (p-q)+ \Delta {\mathbb M}_{pq}$ where ${\mathbb M}^{(0)}$ is diagonal as indicated. It was referred to as a pole term and $\Delta {\mathbb M}$ was designated a cut-term, based on the method of calculation we had used \cite{Kabat:2010nm}.
More appropriately, ${\mathbb M}^{(0)}$ was a ``direct term" giving the geometrical optics approximation while the ``diffractive term" $\Delta {\mathbb M}$ captured the effects of diffraction
\cite{thermal}. We then carried out an expansion in powers of the diffractive term, i.e.,
\beqar
\Tr \log ( {\mathbb M}^{(0)} + \Delta {\mathbb M} )
&=& \Tr \log {\mathbb M}^{(0)} + \Tr \bigl( ( {\mathbb M}^{(0)})^{-1} \, \Delta {\mathbb M}\bigr)
\nonumber\\
&& \hskip .2in -{1\over 2} \Tr \bigl( ( {\mathbb M}^{(0)})^{-1} \, \Delta {\mathbb M} \,( {\mathbb M}^{(0)})^{-1} \, \Delta {\mathbb M} \bigr) + \cdots\label{expansion}
\eeqar
Although there is really no parameter controlling this expansion, numerically the higher and higher order diffractive contributions in (\ref{expansion}) were smaller and smaller, and a sensible truncation was possible.

We propose to do a similar expansion here. The propagators in (\ref{PHP5k}, \ref{PHP5m}), as we have presented them, naturally show this expansion. Likewise, $N_{2L\xi} = - \Delta{\mathbb M}_L$, $N_{2R\xi} = - \Delta {\mathbb M}_R$ are to be considered as of the first order in the diffractive effect. With this understanding, we can now proceed to the individual terms in
(\ref{PHP5r}).

\subsubsection*{\underline{The bulk contribution}}

The bulk contribution to the free energy is
\beq
F_{bulk}= {1\over 2} L_3 (W_1 + W_2) \int {d^3k \over (2\pi )^3} \log \left(
1- e^{- 2 b \omega}\right) \label{PHP5s} 
\eeq
where $W_1$ is the width of the aperture, $W_2$ is the width of the half-plate on the right,
with $L_2 = W_1 + W_2$. 

\subsubsection*{\underline{The contribution from fields on the left plate (plate I)}}

For $F_I$ which is the contribution from the left plate, we have
\begin{align}
F_{I} & = {1\over 2} \Tr \log ( \omega_k \coth b\omega_k ~+ \kappa_I )
~-~ (b \rightarrow \infty ~{\rm limit})\nonumber\\
&= {1\over 2} L_3 (W_1+ W_2) \int {d^3k \over (2\pi )^3} \log \left[ {\omega_k \coth b \omega_k + \kappa_I \over \omega_k + \kappa_I }\right]
\label{PHP5t}
\end{align}
We have also carried out the renormalization by subtracting the
$b \rightarrow \infty$ limit.

\subsubsection*{\underline{The contribution from fields on the left side of the half-plate}}

The term $F_{2L}$ corresponds to ${\half} \Tr \log {\mathbb M}_L$ and can be simplified as
\begin{align}
F_{2L} &= {1\over 2} L_3 W_2 \int {d^2k \over (2\pi)^2} {dp \over \pi} 
\log\left( H_p + \kappa_L \right)
+ {1\over 2} L_3 \int {d^2 k \over (2\pi )^2} \Tr \log \left[
1+ {1\over 2 (H_p + \kappa_L )}  \,\Delta{\mathbb M}_L\right] \nonumber\\
&\hskip .3in ~-(b \rightarrow \infty ~{\rm limit})\nonumber\\
&= F_{2L}^{(0)} + F_{2L}^{(1)} + F_{2L}^{(2)} + \cdots
\label{PHP31}
\end{align}
where we have expanded the second term in a series in $\Delta{\mathbb M}_L$. In expanding the last term,
we get the trace of products of $(1/ 2 (H_p +\kappa_L))\,\Delta {\mathbb M}_L$. The
``$\Tr$" stands for setting the labels
$p=q$ for the two end terms and integrating with $dp/\pi$.
The first few terms in the expansion are
\begin{align}
F_{2L}^{(0)}& =  {1\over 2} L_3 W_2 \int {d^2k \over (2\pi)^2} {dp \over \pi} 
\log\left( {H_p + \kappa_L\over \omega_p + \kappa_L } \right)\label{PHP35a}\\
F^{(1)}_{2L} & = {L_3\over 2}  \int {d^2 k \over (2\pi )^2} {d p \over \pi} \,  
 {1\over 2 (H_p + \kappa_L )}  \,(\Delta{\mathbb M}_L)_{pp} ~-~ (b\rightarrow \infty ~{\rm limit})\label{PHP35b}\\
 F^{(2)}_{2L} &= - {L_3\over 4} \int {d^2k \over (2\pi )^2} {d p \over \pi} {d q\over \pi}\,
  {1\over 2 (H_p + \kappa_L) } (\Delta{\mathbb M}_L)_{pq} {1\over 2 (H_q + \kappa_L)} (\Delta{\mathbb M}_L)_{qp}~-~ (b\rightarrow \infty ~{\rm limit})\label{PHP35c}
\end{align}

\subsubsection*{\underline{The contribution from fields on the right side of the half-plate}}

The term $F_{2R}$ corresponds to ${\half} \Tr \log {\mathbb M}_R$;
it has no $b$-dependence and hence it is eliminated by renormalization.

\subsubsection*{\underline{The contribution from fields on the aperture}}

The next term of interest is the contribution from the $\xi$-integration. The $\xi\,\xi$ terms in
(\ref{PHP5d2}) can be collected together as
\beqar
S_{\xi\xi}&=& {1\over 2} \int_{\vk, p,q} \xi(-\vk, p) \left[ {\mathbb M}_\xi (p,q) -
\int_{p', q'} N_{2L\xi}(\vk,p',p) \, G_{2L}(\vk, p',q')\,  N_{2L\xi} (\vk, q',q)\right. \nonumber\\
&& \hskip 1in 
\left. - \int_{p', q'} N_{2R\xi}(\vk,p',p) \, G_{2R}(\vk, p',q')\,  N_{2R\xi} (\vk, q',q) \right] \, \xi (\vk, q)
\nonumber\\
&=&  {1\over 2} \left[ \int_{\vk,p} \xi(-\vk, p) \, 2 \,(H_p +\omega_p) \, \xi (\vk, p)
+ \int_{\vk,p,q} \xi (-\vk, p) (\Delta {\widetilde {\mathbb M}_\xi})_{pq} \, \xi(\vk,q)
\right]
\label{PHP55a}
\eeqar
where
\beqar
(\Delta {\widetilde {\mathbb M}_\xi})_{pq}
&=& (\Delta {{\mathbb M}_\xi})_{pq} - \int_{p', q'} N_{2L\xi}(\vk,p',p) \, G_{2L}(\vk, p',q')\,  N_{2L\xi} (\vk, q',q) \nonumber\\
&&\hskip .3in -\int_{p', q'} N_{2R\xi}(\vk,p',p) \, G_{2R}(\vk, p',q')\,  N_{2R\xi} (\vk, q',q)
\nonumber\\
&\equiv& (\Delta {\widetilde {\mathbb M}_\xi}^{(1)}) _{pq}
+ (\Delta {\widetilde {\mathbb M}_\xi})^{(2)}_{pq} +\cdots\label{PHP55b}\\
(\Delta {\widetilde {\mathbb M}_\xi})^{(1)}_{pq} &=&
(\Delta {{\mathbb M}_\xi})_{pq}\label{PHP55c}\\
(\Delta {\widetilde {\mathbb M}_\xi})^{(2)}_{pq} &=& - \int_{p'} N_{2L\xi}(\vk,p',p) {1\over 2 (H_{p'} + \kappa_L)}
N_{2L\xi}(\vk, p', q) \nonumber\\
&&\hskip .1in- \int_{p'} N_{2R\xi}(\vk,p',p) {1\over 2 (\omega_{p'} + \kappa_R)}
N_{2R\xi}(\vk, p', q) \label{PHP55d}
\eeqar
For the contribution to free energy upon integration over the $\xi$'s we find
\beqar
F_{\xi} &=& {1\over 2} L_3 W_1 \int {d^2k \over (2\pi )^2}  {d p \over \pi} 
\log (H_p +\omega_p ) + {1\over 2} L_3 \int {d^2k \over (2\pi )^2} \Tr \log \left( 1+
 {1\over 2 (H_p +\omega_p)} ( {\Delta \widetilde{\mathbb M}_\xi} ) \right)\nonumber\\
&&\hskip .5in  ~-~ (b \rightarrow \infty ~{\rm limit})
\label{PHP30}
\eeqar
We can simplify this further by noting that
\beq
\left[{H_p +\omega_p  \over (H_p+\omega_p)_{b \rightarrow \infty } }\right] = 
\left( {1\over 1- e^{- 2 b \omega_p}}\right)\, \left( {\omega_p + \kappa_I \over \omega_p \coth b \omega_p + \kappa_I }\right)
\label{PHP32}
\eeq
The term $F_{\xi}$ thus splits up as
\begin{align}
F_\xi &= F^{(0)}_\xi + F^{(1)}_{\xi} + F^{(2)}_{\xi} + \cdots \nonumber\\
F^{(0)}_\xi  &= - {1\over 2} L_3 W_1 \int {d^3k \over (2\pi )^3} \log \left( 1- e^{-2 b \omega_k}\right)
- {1\over 2} L_3 W_1 \int {d^3k \over (2\pi )^3} \log \left[ {\omega_k \coth b\omega_k + \kappa_I
\over \omega_k + \kappa_I}\right]\label{PHP33a}\\
F^{(1)}_\xi& =  {L_3\over 2} \int {d^2k \over (2\pi )^2} {d p \over \pi}{1\over 2 (H_p +\omega_p) } ({\Delta {\mathbb M}_\xi})_{pp} ~-~ (b\rightarrow \infty ~{\rm limit})\label{PHP33b}\\
F^{(2)}_{\xi}&= - {L_3\over 4} \int {d^2k \over (2\pi )^2} {d p \over \pi} {d q\over \pi} \left[
{1\over 2 (H_p+\omega_p)} (\Delta {\mathbb M}_\xi)_{pq} {1\over 2 (H_p+\omega_p)}( \Delta{\mathbb M})_{qp} 
\right]\nonumber\\
&\hskip .3in + {L_3\over 2} \int {d^2k \over (2\pi )^2} {d p \over \pi}{1\over 2 (H_p +\omega_p) } ({\Delta \widetilde{\mathbb M}_\xi})^{(2)}_{pp}
~-~ (b\rightarrow \infty ~{\rm limit})
 \label{PHP33c}
\end{align}
Notice that the terms in (\ref{PHP33a}) cancel against similar terms in $F_{bulk}$ and $F_{I}$.

\subsubsection*{\underline{The contribution from the field at the edge (field $\rho$)}}

The contributions we have evaluated so far correspond to the fields
$\phi_{2L}$, $\phi_{2R}$ and $\xi$ which vanish at the edge of the half-plate. If the parameter $c_0 \rightarrow \infty$, these are the only terms we have.
For finite values of $c_0$, we have the contribution from $\rho$ as well. To evaluate this term, we need the kernel for the $\rho \rho$-term in the boundary action. In addition to the $\rho\rho$-term
manifestly displayed in (\ref{PHP5d2}), there will be additional terms from the integration over
$\xi$ because of the $\xi$-$\rho$ terms in (\ref{PHP5d2}). For this, first define the 
propagator for $\xi$ given by (\ref{PHP55a}),
\beqar
G_\xi (\vk) &=& \pi \, \delta(p,q) \,{1\over 2 (H_p + \omega_p)}
- {1\over 2 (H_p + \omega_p)} (\Delta\widetilde{\mathbb M}_\xi)_{pq} \, {1\over 2 (H_q + \omega_q)}\nonumber\\
&&+ \int_0^\infty {dp'\over \pi}  {1\over 2 (H_p +\omega_p)} (\Delta\widetilde{\mathbb M}_\xi)_{pp'} \, {1\over 2 (H_p' + \omega_{p'})}
 (\Delta\widetilde{\mathbb M}_\xi)_{p' q} \, {1\over 2 (H_q + \omega_q)}
+\cdots\label{rho1}
\eeqar
The integration over the $\xi$'s thus leads to the $\rho$-dependent terms
\beq
S_{\rho} = {1 \over 2}   \int_{\vk} \rho(-\vk )
\left[ {\mathbb M}_\rho (\vk) + \Delta{\mathbb M}_\rho \right]\, \rho (\vk)
\label{rho2}
\eeq
where
\beqar
\Delta {\mathbb M}_\rho &=&
 - \int_{p,q}N_{2L\rho}(\vk, p) G_{2L}(\vk, p, q) N_{2L\rho}(\vk, q)
 - \int_{p,q} N_{2R\rho}(\vk, p) G_{2R}(\vk, p, q) N_{2R\rho}(\vk, q) \nonumber\\
&& \hskip .3in  - \int_{p,q}{\widetilde N}_{\rho \xi} (-\vk,p) G_\xi (\vk,p,q) {\widetilde N}_{\rho\xi} (\vk,q)
\label{rho3}\\
\widetilde{N}_{\rho\xi} &=& N_{\rho\xi} - \int_{p',q'} N_{2L\xi} (\vk, p', p) G_{2L}(\vk, p', q')
N_{2L\rho} (\vk, q') \nonumber\\
&&\hskip .25in - \int_{p',q'} N_{2R\xi} (\vk, p', p) G_{2R}(\vk, p', q')
N_{2R\rho} (\vk, q')
\label{rho4}
\eeqar
The quantity $\Delta {\mathbb M}_\rho$ naturally has an expansion which follows from the expansion of the propagators $G_{2L}$, $G_{2R}$ and $G_\xi$. Notice that we start off with at least two powers of $N_{2L\xi}$, $N_{2R\xi}$ or $N_{\rho\xi}$, so that there is no
first order term. The expansion for $\Delta {\mathbb M}_\rho$ is thus
\beqar
\Delta {\mathbb M}_\rho &=& \Delta {\mathbb M}^{(2)}_\rho
+ \Delta {\mathbb M}^{(3)}_\rho + \cdots\nonumber\\
\Delta {\mathbb M}^{(2)}_\rho &=&  - \int_{p}N_{2L\rho}(\vk, p) {1\over 2(H_p +\kappa_L)} N_{2L\rho}(\vk, p)
 - \int_{p} N_{2R\rho}(\vk, p) {1\over 2(\omega_p +\kappa_R)} N_{2R\rho}(\vk, p) \nonumber\\
&& \hskip .3in  - \int_{p}{ N}_{\rho \xi} (-\vk,p) {1\over 2(H_p +\omega_p)} { N}_{\rho\xi} (\vk, p)\label{rho5}
\eeqar
All terms in these expressions depend on $\vk$; we have not indicated this explicitly in the arguments of various functions to avoid too much clutter.
Once the integrations over $p, q$, etc. are done, 
${\mathbb M}_\rho + \Delta {\mathbb M}_\rho$ is only a function of
$\vk$, so that the free energy resulting from (\ref{rho2}) can be written as
\beq
F_\rho = {L_3 \over 2} \int {d^2k \over (2\pi)^2}\,
\log \left(  {\mathbb M}_\rho + \Delta{\mathbb M}_\rho\right)
\label{rho7}
\eeq
Thus approximations to this term will be of the form
\beqar
F^{(0)}_\rho &=& {L_3 \over 2} \int {d^2k \over (2\pi)^2}\,
\log \left(  {\mathbb M}_\rho \right) - (b \rightarrow \infty ~{\rm limit})\nonumber\\
F^{(0)}_\rho + F^{(2)}_\rho&=& {L_3 \over 2} \int {d^2k \over (2\pi)^2}\,
\log \left(  {\mathbb M}_\rho + \Delta {\mathbb M}^{(2)}_\rho \right)  - (b \rightarrow \infty ~{\rm limit})
 \label{rho8}
\eeqar

\subsection{Numerical estimates}

The terms proportional to $W_1$, the width of the aperture, cancel out between
(\ref{PHP5s}), (\ref{PHP5t}) and (\ref{PHP33a}). Thus we are left with bulk terms which only involve the facing area $L_3 W_2$ between the two plates. This is given by
part of $F_{bulk}$ in (\ref{PHP5s}), part of $F_I$ in (\ref{PHP5t}) and
$F^{(0)}_{2L}$ in (\ref{PHP35a}) and is equal to
\beq
F_{area} = {L_3 W_2\over 2}
\int {d^3k \over (2\pi)^3} \left[
\log \left( 1 - e^{-2b\omega_k}\right) + \log\left( {\omega_k \coth b\omega_k + \kappa_I \over
\omega_k + \kappa_I}\right) + \log \left( {H_k + \kappa_L \over \omega_k +\kappa_L}\right)
\right]\label{energy1}
\eeq

There are also many edge-dependent terms to consider. The fields
$\xi$, $\phi_{2L}$ vanish at the edge $x_2 =0$. For an aperture 
and a half-plate of finite widths $W_1$ and $W_2$, we should use discrete modes
to evaluate the free energy, with a sum replacing the integral over $p$ or $k_2$.
In the limit of large $W_1$, $W_2$, the sum over the discrete modes can be approximated by an integral, as in an Euler-Maclaurin summation formula,
which gives the area-dependent term
in (\ref{energy1}). But there is also a subdominant edge-dependent term in the summation 
formula \cite{Kabat:2010nm}; we will denote this term by $F^{(0)}_{edge}$.

The remaining edge-dependent terms in the energy are given by
$F^{(1)}_{2L} + F^{(2)}_{2L} +\cdots$ from (\ref{PHP35b}), (\ref{PHP35c}), etc.,
$F^{(1)}_\xi + F^{(2)}_\xi +\cdots$ from (\ref{PHP33b}), (\ref{PHP33c}), etc., and
$F_\rho$ from (\ref{rho8}).
All these quantities depend on the parameters $\kappa_I$, $\kappa_L$, $\kappa_R$,
$c_0$, $c_{2L}$ and $c_{2R}$. We will now proceed to the numerical estimate of these integrals for some choices of these parameters. 

\subsubsection*{\underline{The area-dependent contribution}}

\begin{table}[b!]
\begin{center}
\setlength{\extrarowheight}{2pt}
\caption{The area-dependent contributions ${\cal E}_{area}$ for different boundary conditions}
\vskip .1in
\begin{tabular}{| c | c c c c c c| }
\hline
$m$&0&1&2&3&4&5\\
\hline
&&&&&&\\
~~${\cal E}_{RD}(\infty , m)={\cal E}_{DR}(m,\infty )$~~&~~~0.875~~~&0.092&-0.212&-0.381&-0.489&-0.564\\
~~${\cal E}_{RR}(m, m)$~~&~-1~&-0.089&0.115&-0.200&-0.283&-0.354\\
&&&&&&\\
\hline
$m$&6&7&8&9&10&$\infty$\\
\hline
&&&&&&\\
~~${\cal E}_{RD}(\infty , m)={\cal E}_{DR}(m,\infty )$~~&-0.620&-0.663&-0.698&-0.725&-0.749&-1\\
~~${\cal E}_{RR}(m, m)$~~&-0.414&-0.464&-0.507&-0.544&-0.575&-1\\
&&&&&&\\
\hline
\end{tabular}
\label{tab1}
\end{center}
\end{table}
\begin{figure}[!ht]
\begin{center}
\scalebox{.30}{\includegraphics{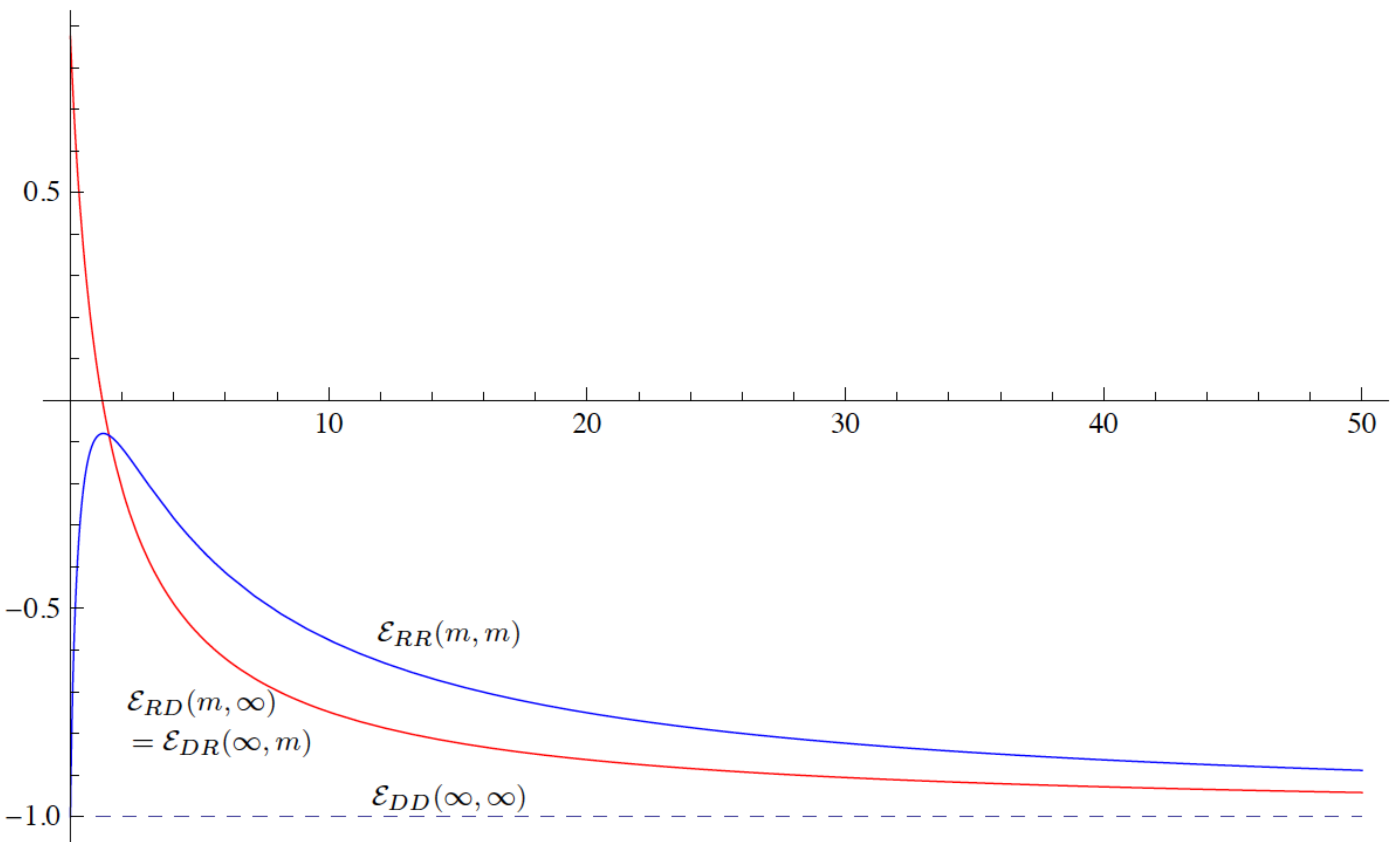}}
\caption{The area-dependent contribution to energy, ${\cal E}_{area}$, for different boundary conditions,
as a function of $m$, showing asymptotic approach to the Dirichlet-Dirichlet case.}
\label{pic4}
\end{center}
\end{figure}
The area-dependent part of the Casimir energy (\ref{energy1}), from scaling out $b$, is of the form
\beq
F_{area} = {L_3 W_2 \over b^3} \,{\pi^2 \over 1440}\,~ {\cal E}_{area}(b\kappa_I, b\kappa_L )
\label{energy2}
\eeq
We will consider Dirichlet condition for the left plate ($\kappa_I \rightarrow \infty$), Robin condition for the right (with finite $b\kappa_L =m$), or the other way ($\kappa_L \rightarrow \infty, \, b\kappa_I =m$ finite), and then Robin condition for both (with $b\kappa_I =
b\kappa_L = m$).  The values for these cases are given in Table \ref{tab1}.
The Dirichlet-Robin and Robin-Dirichlet cases give the same area-dependent term for the energy. 
The Robin-Robin case with $m =0$ corresponds to Neumann condition on both plates.
The limit of $m \rightarrow \infty$ should correspond to the case with Dirichlet conditions on both plates which is, of course, well known.
As expected, and as Fig.\,\ref{pic4} shows, the values found here approach that value (${\cal E} = -1$)
asymptotically.

\subsubsection*{\underline{The edge-contributions (without $\rho$)}}

The edge-dependent contributions from $F_{2L}$ and $F_\xi$ (including
$F^{(0)}_{edge}$) are of the form
\beq
F_{edge} = {L_3 \over b^2} \,~ {\cal E}_{edge} (b\kappa_I, b \kappa_L, b\kappa_R)
\label{energy3}
\eeq
We consider the cases of  Robin-Dirichlet ($ b \kappa_I$ = m = finite value, $ \kappa_L , \kappa_R \rightarrow
\infty$), Dirichlet-Robin ($\kappa_I \rightarrow \infty$,
$b \kappa_L = b \kappa_R$= m = finite value) and Robin-Robin ($b \kappa_I = b \kappa_L =  b \kappa_R$ =  m=finite value).
The values ${\cal E}_{edge}$ for a range of $m$ are given in Tables \ref{tab2}, \ref{tab2a}, \ref{tab2b}.

\begin{figure}[ht!]
\begin{center}
\vskip .3in

\scalebox{.38}{\includegraphics{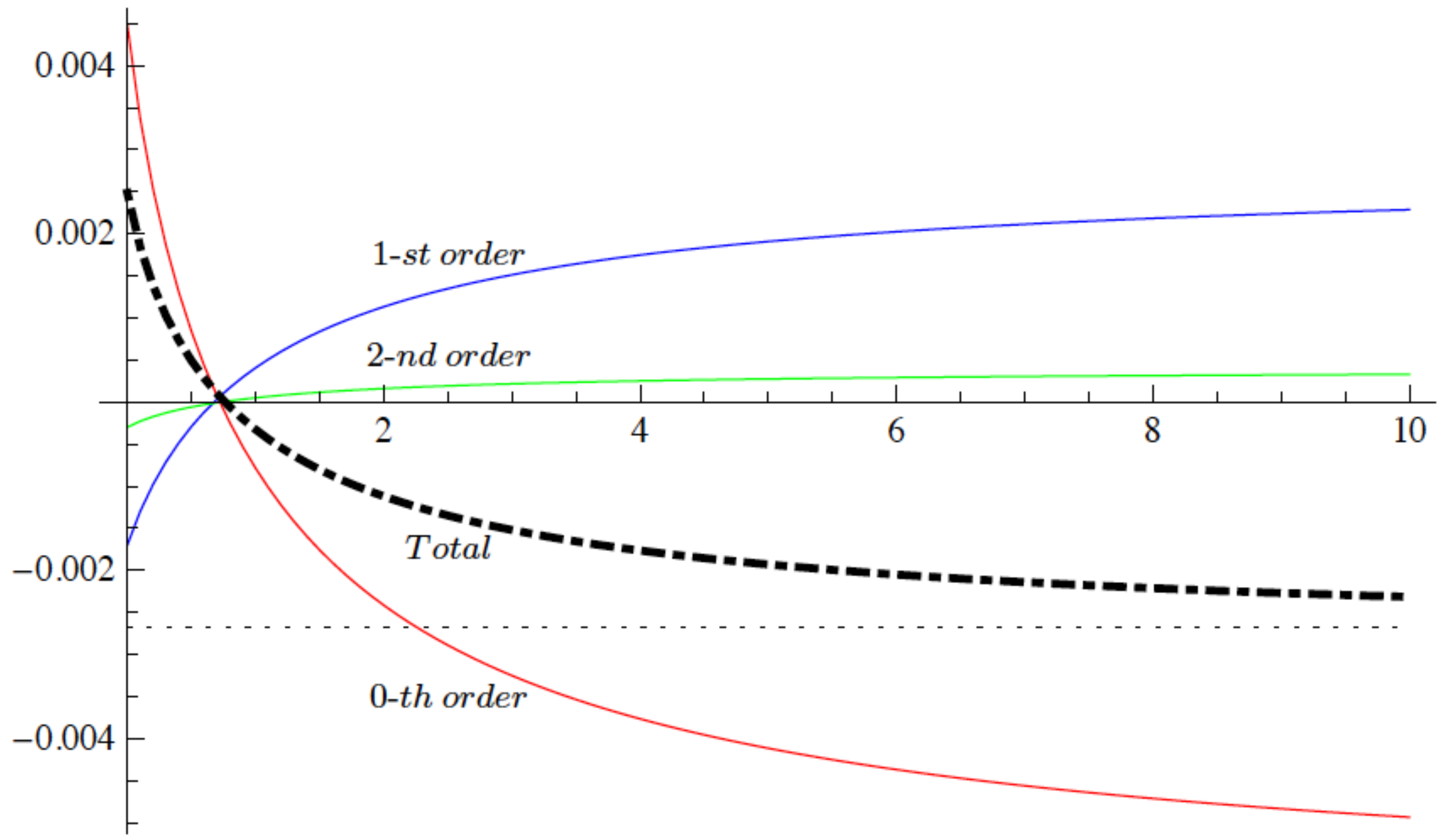}}
\caption{The edge-dependent contributions from $F_{2L}$ and $F_\xi$ for Robin-Dirichlet case, 
${\cal E}_{edge, RD}(m, \infty, \infty)$.}
\label{pic6}
\end{center}
\end{figure}

\begin{figure}[ht!]
\begin{center}
\vskip .3in

\scalebox{.38}{\includegraphics{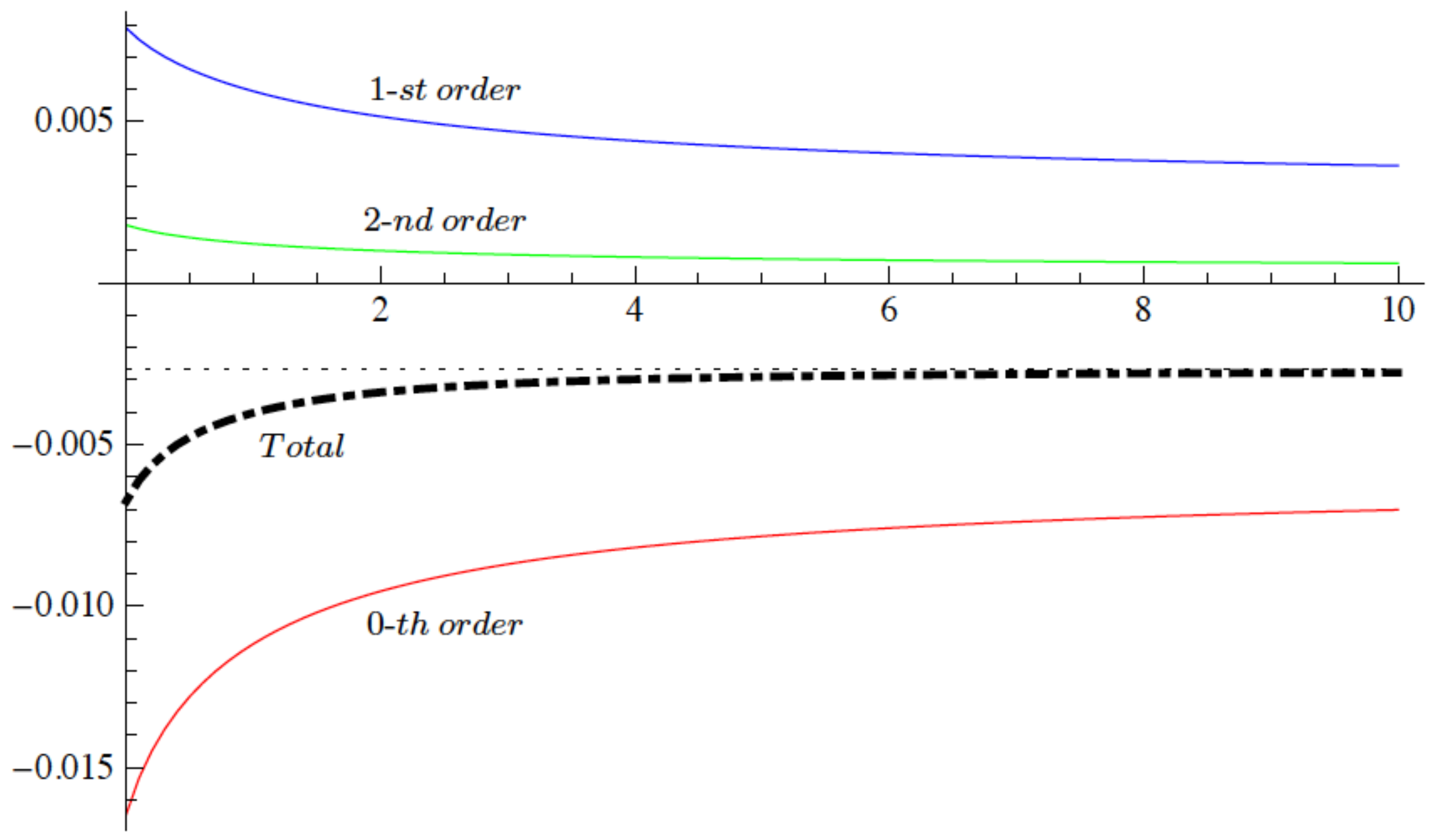}}
\caption{The edge-dependent contributions from $F_{2L}$ and $F_\xi$ for Dirichlet-Robin case, 
${\cal E}_{edge,DR}(\infty, m, m)$.}
\label{pic5}
\end{center}
\end{figure}

\begin{figure}[ht!]
\begin{center}
\scalebox{.38}{\includegraphics{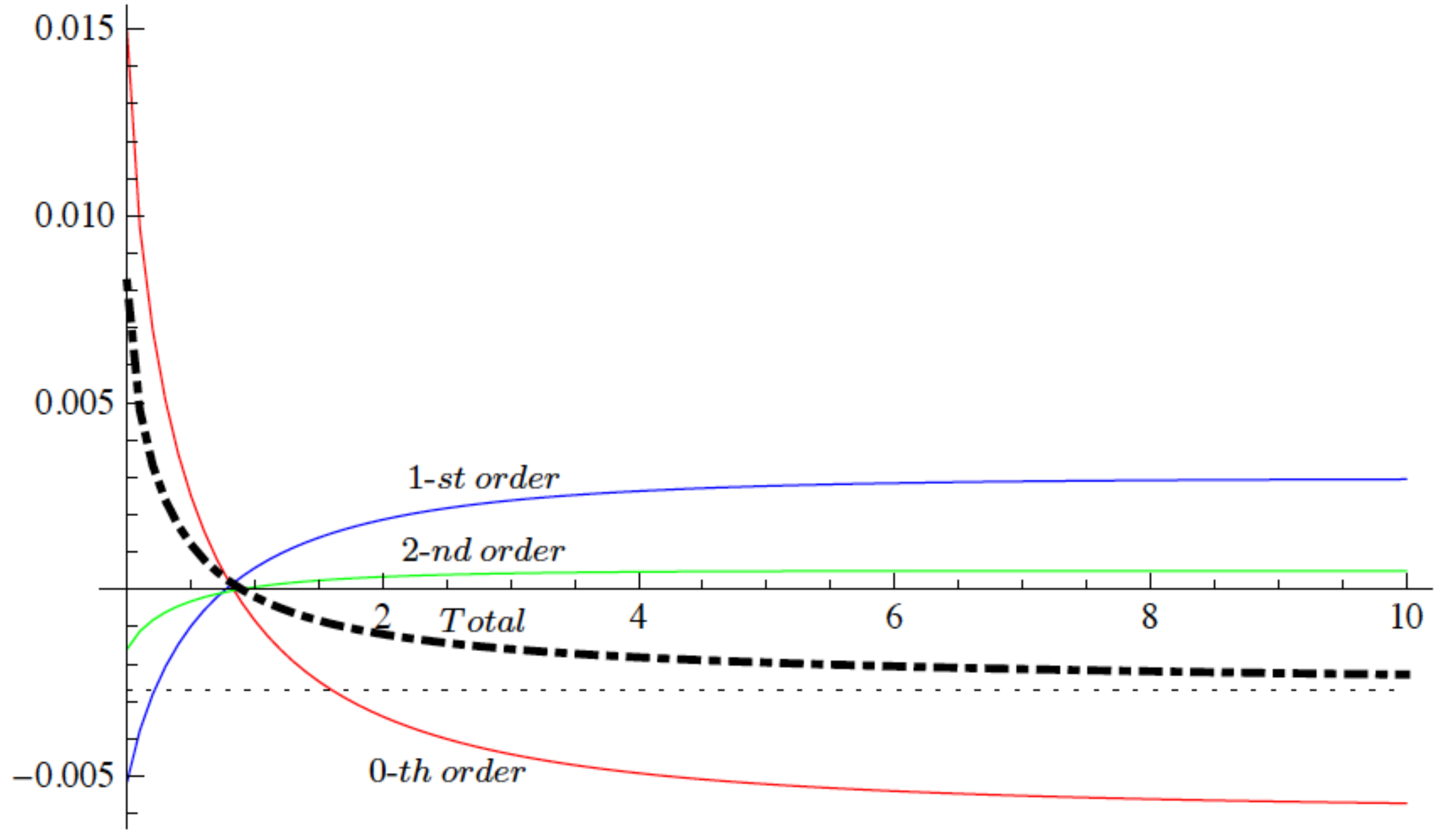}}
\caption{The edge-dependent contributions from $F_{2L}$ and $F_\xi$ for Robin-Robin case, 
${\cal E}_{edge, RR}(m, m, m)$.}
\label{pic7}
\end{center}
\end{figure}

We have calculated these to the second order in the expansion of the $\Tr \log$ terms, as in
(\ref{PHP35a}-\ref{PHP35c}), (\ref{PHP33a}-\ref{PHP33c}). These do not include the contribution from the $\rho$-integration which will be given separately. We also graphically display
the leading order term, the first order term, the second order term and the total edge contribution to this order for the three cases in Figs.\,\ref{pic6}, \ref{pic5}, \ref{pic7}.
The dashed line is the total contribution, up to second order, and the dotted line is the Dirichlet-Dirichlet limit.

\vfill\eject

\begin{table}[!htbp]
\begin{center}
\setlength{\extrarowheight}{2pt}
\caption{The edge contribution to the Casimir energy, Robin-Dirichlet case, without $F_\rho$}
\vskip .1in
\begin{tabular}{| c | c c c c c c| }
\hline
$m$&0&1&2&3&4&5\\
\hline
&&&&&&\\
${\cal E}_{edge,RD}^0(m, \infty , \infty)$&0.00448&-0.00079&-0.00241&-0.00325&-0.00377&-0.00412\\
${\cal E}_{edge,RD}^1(m, \infty, \infty)$&-0.00030&0.00005&0.00016&0.00022&0.00025&0.00028\\
${\cal E}_{edge,RD}^2(m,\infty, \infty )$&0.00180&0.00121&0.00100&0.00088&0.00081&0.00075\\
${\cal E}_{edge,RD}^{total}(m, \infty , \infty)$&0.00249&-0.00033&-0.00112&-0.00152&-0.00176&-0.00193\\
&&&&&&\\
\hline
$m$&6&7&8&9&10&$\infty$\\
\hline
&&&&&&\\
${\cal E}_{edge,RD}^0(\infty , m, m)$&-0.00439&-0.00456&-0.00471&-0.00483&-0.00493&-0.00598\\
${\cal E}_{edge,RD}^1(\infty, m,m)$&0.00203&0.00212&0.00219&0.00224&0.00229&0.00277\\
${\cal E}_{edge,RD}^2(\infty , m, m)$&0.00029&0.00031&0.00032&0.00032&0.00033&0.00040\\
${\cal E}_{edge,RD}^{total}(\infty , m, m)$&-0.00205&-0.00214&-0.00221&-0.00227&-0.00232&-0.00280\\
&&&&&&\\
\hline
\end{tabular}
\label{tab2}
\end{center}
\end{table}

\vfill\eject

\begin{table}[!htbp]
\begin{center}
\setlength{\extrarowheight}{2pt}
\caption{The edge contribution to the Casimir energy, Dirichlet-Robin case, without $F_\rho$}
\vskip .1in
\begin{tabular}{| c | c c c c c c| }
\hline
$m$&0&1&2&3&4&5\\
\hline
&&&&&&\\
${\cal E}_{edge,DR}^0(\infty , m, m)$&-0.01644&-0.01117&-0.00954&-0.00871&-0.00819&-0.00784\\
${\cal E}_{edge,DR}^1(\infty, m,m)$&0.00790&0.00594&0.00516&0.00471&0.00441&0.00419\\
${\cal E}_{edge,DR}^2(\infty,m , m, )$&0.00180&0.00121&0.00100&0.00088&0.00081&0.00075\\
${\cal E}_{edge,DR}^{total}(\infty , m, m)$&-0.00675&-0.00402&-0.00338&-0.00312&-0.00298&-0.00290\\
&&&&&&\\
\hline
$m$&6&7&8&9&10&$\infty$\\
\hline
&&&&&&\\
${\cal E}_{edge,DR}^0(\infty , m, m)$&-0.00759&-0.00740&-0.00725&-0.00712&-0.00702&-0.00598\\
${\cal E}_{edge,DR}^1(\infty, m,m)$&0.00403&0.00390&0.00379&0.00371&0.00364&0.00277\\
${\cal E}_{edge,DR}^2(\infty , m, m)$&0.00071&0.00068&0.00065&0.00063&0.00061&0.00040\\
${\cal E}_{edge,DR}^{total}(\infty , m, m)$&-0.00285&-0.00282&-0.00280&-0.00278&-0.00277&-0.00280\\
&&&&&&\\
\hline
\end{tabular}
\label{tab2a}
\end{center}
\end{table}

\vfill\eject

\begin{table}[!ht]
\begin{center}
\setlength{\extrarowheight}{2pt}
\caption{The edge contribution to the Casimir energy, Robin-Robin case, without $F_\rho$}
\vskip .1in
\begin{tabular}{| c | c c c c c c| }
\hline
$m$&0&1&2&3&4&5\\
\hline
&&&&&&\\
${\cal E}_{edge,RR}^0(m, m, m)$&0.01495&-0.00084&-0.00340&-0.00441&-0.00492&-0.00521\\
${\cal E}_{edge,RR}^1(m, m,m)$&-0.00515&0.00058&0.00187&0.00238&0.00263&0.00277\\
${\cal E}_{edge,RR}^2(m, m , m )$&-0.00159&0.00006&0.00034&0.00043&0.00047&0.00049\\
${\cal E}_{edge,RR}^{total}(m, m, m)$&0.00820&-0.00020&-0.00120&-0.00160&-0.00182&-0.00196\\
&&&&&&\\
\hline
$m$&6&7&8&9&10&$\infty$\\
\hline
&&&&&&\\
${\cal E}_{edge,RR}^0(m, m, m)$&-0.00540&-0.00552&-0.00561&-0.00568&-0.00573&-0.00598\\
${\cal E}_{edge,RR}^1(m, m,m)$&0.00285&0.00289&0.00292&0.00294&0.00295&0.00277\\
${\cal E}_{edge,RR}^2(m, m, m)$&0.00049&0.00050&0.00050&0.00050&0.00050&0.00040\\
${\cal E}_{edge,RR}^{total}(m, m, m)$&-0.00206&-0.00213&-0.00220&-0.00224&-0.00228&-0.00280\\
&&&&&&\\
\hline
\end{tabular}
\label{tab2b}
\end{center}
\end{table}
\vskip 0.3in

\vfill\eject

\subsubsection*{\underline{The edge-contributions  from $\rho$}}

\begin{figure}[!t]
\begin{center}


\scalebox{.31}{\includegraphics{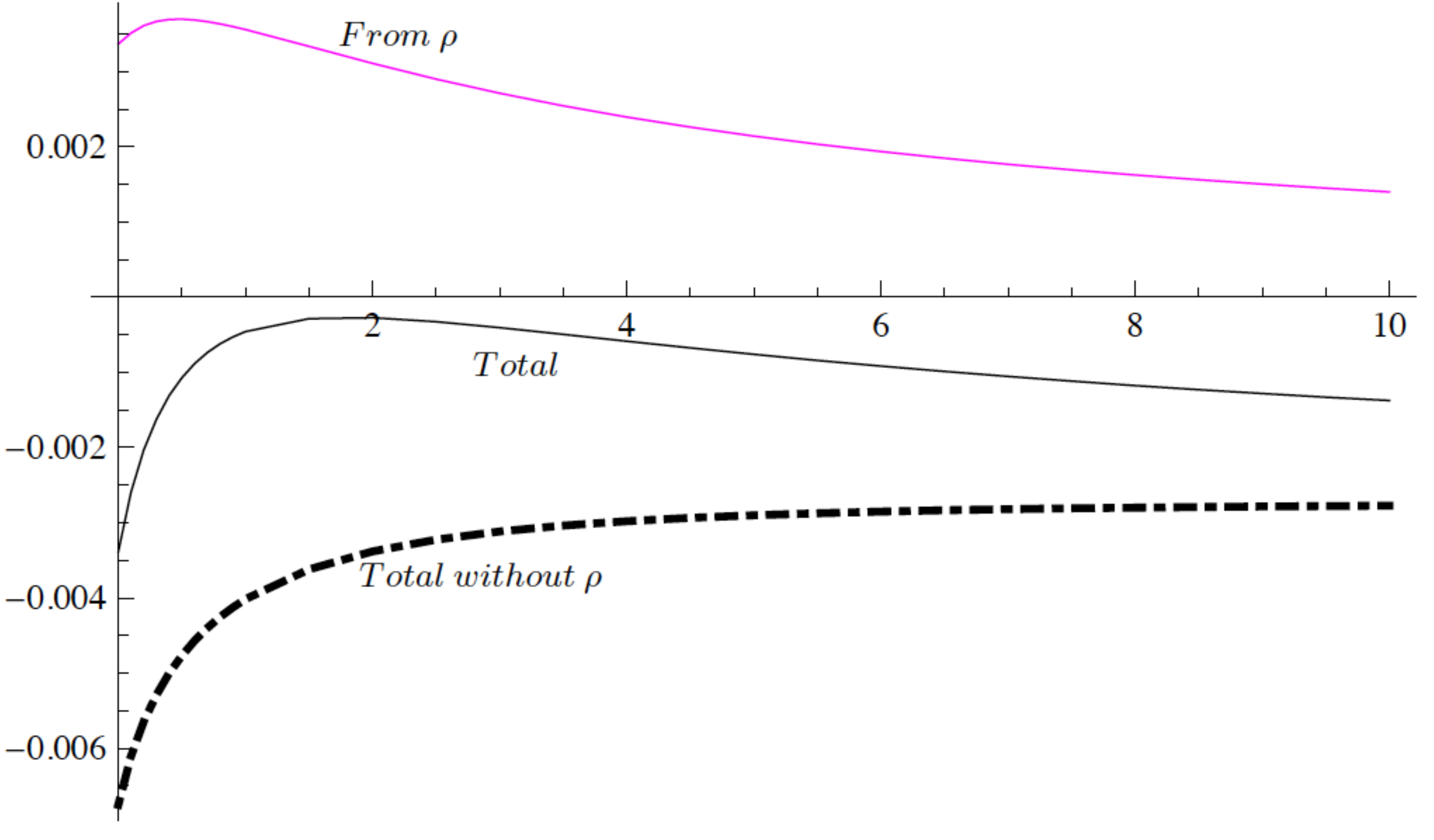}}


\caption{The edge-dependent contributions, including $\rho$, for Dirichlet-Robin case, 
${\cal E}_{edge, DR}(\infty, m, m, 2 m)$.}
\label{pic8}
\end{center}
\end{figure}

\begin{figure}[ht!]
\begin{center}
\scalebox{.31}{\includegraphics{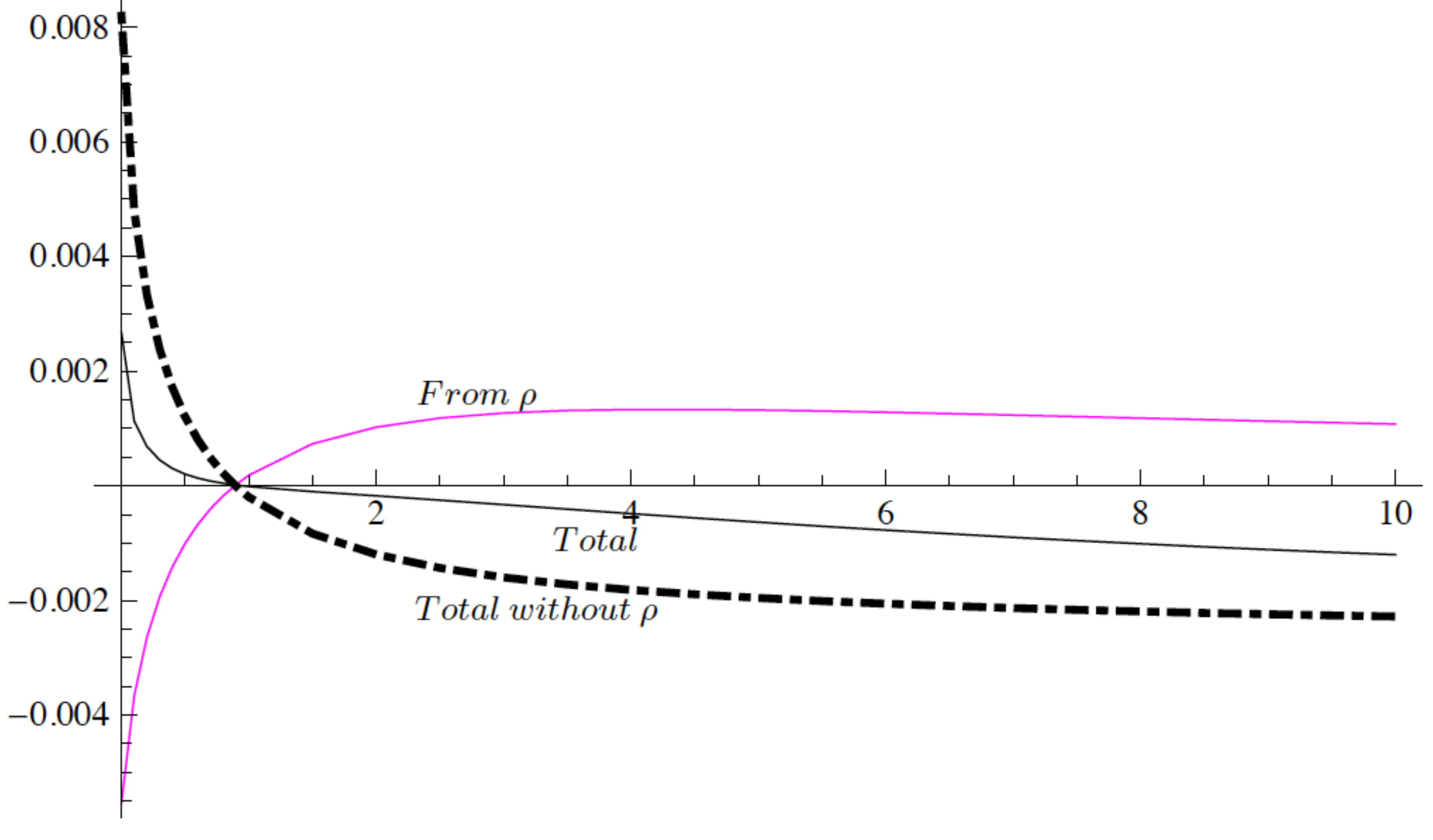}}
\caption{The edge-dependent contributions, including $\rho$ for Robin-Robin case, 
${\cal E}_{edge, RR}(m, m, m, 2 m)$.}
\label{pic9}
\end{center}
\end{figure}

The edge contribution from $\rho$ is zero for the case of $c_0 \rightarrow \infty$. Thus for the 
case of Robin conditions (or anything else) on the left plate and Dirichlet conditions on the half-plate, the results given so far suffice. We will consider the inclusion of the contribution from
$\rho$ for the Dirichlet-Robin case ($\kappa_I \rightarrow \infty$). 
For simplicity, we shall consider the case $c_{2L} = c_{2R} =0$.
We will need to specify
$c_0$ as well. We take it to be $c_0 = (\kappa_L + \kappa_R)$, since this is 
what would be naturally considered if we take $\rho$ to be part of
$\phi_{2L}$ and $\phi_{2R}$. (There is no compelling reason for this choice; it is one case
worth considering and easy enough to calculate.) Thus the contribution from $\rho$ is of the form
\beq
F_{\rho} = {L_3 \over b^2} \,~{\cal E}_\rho ( b\kappa_I, b\kappa_L, b \kappa_R, bc_0)
\label{energy4}
\eeq
We will calculate this to the second order as well, which means that we will evaluate
the expression in (\ref{rho8}). These values, for a range of parameters, are given in 
Table \ref{tab5}. We also show the contribution from $\rho$ as compared to the other edge
contributions as a function of the parameter $m$ in graphs Figs.\,\ref{pic8}, \ref{pic9}.
\begin{table}[!htbp]
\begin{center}
\setlength{\extrarowheight}{2pt}
\caption{The edge contribution from $\rho$ to the Casimir energy}
\vskip .1in
\begin{tabular}{| c | c c c c c c| }
\hline
$m$&0&1&2&3&4&5\\
\hline
&&&&&&\\
${\cal E}_{\rho, DR}(\infty , m, m, 2m)$&
0.00336&0.00356&0.00311&0.00271&0.00239&0.00214\\
${\cal E}_{\rho, RR}(m , m, m, 2m)$
&-0.00553&0.00185&0.00102&0.00127&0.00133&0.00132\\
&&&&&&\\
\hline
$m$&6&7&8&9&10&$\infty$\\
\hline
&&&&&&\\
${\cal E}_{\rho, DR}(\infty , m, m, 2m)$&
0.00193&0.00177&0.00162&0.00150&0.00140&0\\
${\cal E}_{\rho, RR}(m , m, m, 2m)$&
0.00128&0.00123&0.00118&0.00113&0.00108&0\\
&&&&&&\\
\hline
\end{tabular}
\label{tab5}
\end{center}
\end{table}


\section{Discussion}

We have generalized our previous formulation for the calculation of diffractive effects in Casimir energy in the case of boundaries with edges and apertures to include general boundary conditions. As a specific example we have analyzed the geometry of two parallel plates, one of which is a semi-infinite so that there are edge effects and diffraction. We have considered a variety of boundary conditions and almost all the results are new, not calculated previously by any method. There are a few special cases for which results using other methods are known and a comparison with the literature is possible.

\begin{table}[!b]
\begin{center}
\setlength{\extrarowheight}{2pt}
\caption{Summary for DD, ND, DN and NN cases}
\vskip .1in
\begin{tabular}{| c | c | c | c | c | }
\hline
&DD&ND&DN&NN\\
\hline
&&&&\\
${\cal E}_{area}$&-1&${7 \over 8}$&${7 \over 8}$&-1\\
${\cal E}^0_{edge}$&~~-${\zeta (3) \over {64 \pi}}=-0.00598$~~&~~${3 \over 4}{\zeta (3) \over {64 \pi}}=0.00448$~~&~~-${11 \over 4} {\zeta (3) \over {64 \pi}}=-0.01645$~~&~~${10 \over 4} {\zeta (3) \over {64 \pi}}= 0.01495$~~\\
${\cal E}^1_{edge}$&0.00277&-0.00170&0.00790&-0.00515\\
${\cal E}^2_{edge}$&0.00040&-0.00030&0.00180&-0.00060\\
\hline
${\cal E}^{total}_{edge}$&-0.00280&0.0048&-0.00675&0.00820\\
&&&&\\
${\cal E}_{edge,\rho}$&0&0&0.00336&-0.00553\\
\hline
&&&&\\
${\cal E}_{edge}$&-0.00280&0.00249&-0.00339&0.00267\\
$(Total)$&&&&\\
\hline
\end{tabular}
\label{tab6}
\end{center}
\end{table}

The first case is where we have two full parallel plates. The Casimir energy for a massless scalar field subject to Robin boundary conditions on two infinitely long parallel plates has been analyzed before in \cite{Robin, cavalcanti}. Of particular interest is \cite{cavalcanti}, where a boundary term was added to impose Robin conditions at the level of Green's functions. The idea is similar to ours although our argument for such a term, which is based on the wave functional interpretation is different and somewhat more general.  If we ignore the diffractive contributions in our calculation and replace the facing area of the plates by the full area our results are in agreement with \cite{Robin, cavalcanti}. (The graph in \cite{cavalcanti} shows ${\cal E}/m^3$ rather than ${\cal E}$ as in our graph.) 

The MIT group has analyzed, using methods of scattering theory, the edge contribution to the Casimir energy in a number of different geometries of boundaries with edges \cite{MIT1, MIT2}. One particular case is exactly the geometry we have but restricted to DD and NN conditions. 
Our results for the DD, ND, DN and NN cases are summarized in Table \ref{tab6}.
These values for the DD and NN cases are in good agreement with the results in 
\cite{Gies, MIT1, MIT2}. In particular, the numerical values obtained from the expressions given in 
\cite{MIT2} are
${\cal E}_{DD}^{edge} = -2.63 \times 10^{-3}$ and
${\cal E}_{NN}^{edge} = 2.97 \times 10^{-3}$. 
In our previous paper \cite{Kabat:2010nm} we calculated the edge contribution up to the 5th order in diffraction, ${\cal E}_{DD}^{edge}=-2.68 \times 10^{-3}$, and the agreement with \cite{MIT2} is even better than is apparent from Table
\ref{tab6}. In making this comparison, it is important to keep in mind that the Neumann boundary condition corresponds to integrating over all boundary fields, including $\rho$ (with $\kappa_I=\kappa_L=\kappa_R=c_0=0$); hence the contribution from $\rho$ must also be taken into account in comparing the values.
We also note that the DD case was initially treated using numerical worldline methods in \cite{Gies}. The value
obtained was ${\cal E}_{DD}^{edge} = -2.62 \times 10^{-3}$; again the later calculations
\cite{Kabat:2010nm, MIT1, MIT2} are consistent with this result.

Another interesting result which emerges from our calculations is that the diffractive effects are always of the opposite sign to the nondiffractive (and leading) contribution. This had been noticed already in our previous work \cite{Kabat:2010nm, thermal, proceedings} where we considered Dirichlet-Dirichlet boundary conditions. This continues to hold with the more general boundary conditions discussed here, as is strikingly clear from the graphs for the edge contributions.
There is presumably a general reason for this, it is a point worth exploring.

Finally, a short remark on the issue of negative eigenvalues: The bulk determinants in our approach are calculated with Dirichlet boundary conditions. 
There are no negative eigenvalues to worry about for this calculation.
However, this issue is not totally eliminated; it has a lingering effect on the boundary action.
Notice that the signs of $\kappa_I$, $\kappa_L$, etc. are important. These parameters occur
in the boundary action and negative values for these can lead to instabilities. For example,
the propagators  in (\ref{PHP5k}, \ref{PHP5m}) clearly display this possibility.
We have to conclude that
acceptable boundary conditions which encode the boundary effects of real material plates must be such that the operators ${\mathcal K}$ have positive eigenvalues.
One may still ask the question whether there is any meaningful physical interpretation for the mathematically acceptable case of negative eigenvalues. This will be discussed in a separate paper \cite{nair-trg}.

\vskip .1in
This research was supported in part by NSF grants PHY-1068172  and PHY-1213380 
and by PSC-CUNY grants.

\section*{Appendix}
In this appendix, we work out some of the simplification of the terms in the boundary action.
Using the mode expansions in (\ref{PHP3}, \ref{PHP3a}, \ref{PHP3b}) and
(\ref{PHP4}), we find
\vskip .1in\noindent
\underline{$\vf_I \,\vf_I$ term}
\begin{align}
{1\over 2} \int \vf_I ( \omega \coth b \omega ~+ \kappa_I ) \, \vf_I
&= {1\over 2} \int {d^3k \over (2\pi )^3} \phi_1(-k) \, {\mathcal M}_1(k) \, \phi_1 (k)\nonumber\\
{\mathcal M}_1(k) &=  \left( \omega_k \coth b \omega_k + \kappa_I \, \right)
\label{PHP6}
\end{align}
\vskip .1in\noindent
\underline{$\vf_I \,\phi_{2L}$ term}
\begin{align}
 \int \vf_I (-\omega\, {\rm csch }\, b \omega ) \phi_{2L}
 &= \int {d^3k \over (2\pi )^3} {d p\over \pi} ~\phi_1 (-k)\, N_{1\,2L}(k, p) \,  \phi_{2L}(\vk, p)
 \nonumber\\
N_{1\,2L}(k, p) &=   {2 p \over p^2 - (k_2 + i\epsilon )^2} ( \omega_K\, {\rm csch} \, b \omega_k )
 \label{PHP7}
\end{align}
\vskip .1in\noindent
\underline{$\vf_I \,\xi$ term}
\begin{align}
 \int \vf_I ( - \omega\, {\rm csch}\, b \omega )\, \xi &=
 \int {d^3k \over (2\pi )^3} {d q \over \pi} ~\phi_1(-k) \, N_{1\xi}(k, q)\,  \xi(\vk, q) 
 \nonumber\\
 N_{1\xi}(k, q)&= {2 q \over q^2 - (k_2 -i\epsilon )^2} (- \omega_k\, {\rm csch} \, b \omega_k )
 \label{PHP7}
\end{align}
\vskip .1in\noindent
\underline{$\xi\,\xi$ term}
\beq
 {1\over 2} \int \xi\, \left(\omega + \omega \coth b \omega \right) \, \xi  = {1\over 2} \int  {d^2k \over (2\pi )^2} {dp \over \pi} {dq \over \pi} ~
 \xi(-\vk, p) \, {\mathcal M}_\xi ( \vk, p, q) \, \xi( \vk, q)\nonumber
 \eeq
 \beq
{\mathcal M}_\xi (\vk, p, q)=  \int_{-\infty}^\infty  {dk_2 \over 2 \pi} {4 p q \over [p^2 - (k_2 - i \epsilon )^2] \, [q^2
  - (k_2 + i \epsilon )^2]} ~\left(\omega_k +  \omega_k \coth b \omega_k  \right)
  \label{PHP10a}
\eeq
\vskip .1in\noindent
\underline{$\phi_{2L}\,\phi_{2L}$ term}
\beq
  {1\over 2} \int \phi_{2L} (\omega \coth b\omega + \kappa_{L} )\, \phi_{2L}
  = {1\over 2} \int {d^2k \over (2\pi )^2} {dp \over \pi} {dq \over \pi} ~ \phi_{2L}(-\vk, p) \, {\mathcal M}_{2L} (\vk, p, q)\,
  \phi_{2L} (\vk, q)\nonumber
  \eeq
  \beq
{\mathcal M}_{2L} (\vk, p, q) = \int_{-\infty}^\infty {dk_2 \over 2 \pi} {4 p q \over [p^2 - (k_2 - i \epsilon )^2] \, [q^2
  - (k_2 + i \epsilon )^2]} ~( \omega_k \coth b \omega_k + \kappa_{L} )
  \label{PHP8}
\eeq
\vskip .1in\noindent
\underline{$\phi_{2R}\,\phi_{2R}$ term}
\begin{align}
{1\over 2} \int \phi_{2R} (\omega + \kappa_{R} ) \phi_{2R}
&= {1\over 2}  \int {d^2k \over (2\pi )^2} {dp \over \pi} {dq \over \pi} ~ \phi_{2R}(-\vk, p) \, {\mathcal M}_{2R} (\vk, p, q)\,
  \phi_{2R} (\vk, q)\nonumber\\
{\mathcal M}_{2R} (\vk, p,q)&=  \int {dk_2 \over 2 \pi} {4 p q \over [p^2 - (k_2 - i \epsilon )^2] \, [q^2
  - (k_2 + i \epsilon )^2]} ~(\omega_k + \kappa_{R} )
  \label{PHP10b}
  \end{align}
\vskip .1in\noindent
\underline{$\xi\,\phi_{2L}$ term}
\begin{align}
  \int \xi \,\,( \omega \coth b \omega )\, \phi_{2L}
  &=  \int  {d^2k \over (2\pi )^2} {dp \over \pi} {dq \over \pi} ~
  \xi (-\vk, p) \, Q_{2L\xi}(\vk, p, q) \, \phi_{2L} (\vk, q) \nonumber\\
Q_{2L\xi}(\vk, p, q)&= - \int_{-\infty}^\infty {dk_2 \over 2 \pi} {4 p q \over [p^2 - (k_2 - i \epsilon )^2] \, [q^2
  - (k_2 - i \epsilon )^2]} ~( \omega_k \coth b \omega_k )
  \label{PHP9a}
\end{align}
\vskip .1in\noindent
\underline{$\xi\,\phi_{2R}$ term}
\begin{align}
  \int \xi \,\,( \omega )\, \phi_{2R}
  &=  \int  {d^2k \over (2\pi )^2} {dp \over \pi} {dq \over \pi} ~
\xi (-\vk, p) \,N_{2R\xi} (\vk, p, q) \, \phi_{2R} (\vk, q) \nonumber\\
N_{2R\xi} (\vk, p, q)&=  - \int {dk_2 \over 2 \pi} {4 p q \over [p^2 - (k_2 - i \epsilon )^2] \, [q^2
  - (k_2 - i \epsilon )^2]} ~ \omega_k 
  \label{PHP9b}
\end{align}

Before taking up the $\rho$-dependent terms, we will consider some simplification of the expressions so far.
We start with ${\mathcal M}_\xi$ and use the integral
representation
\beq
\omega + \omega \coth b\omega
= \int_{-\infty}^\infty {d \lambda \over \pi} {\omega^2 \over \lambda^2 + \omega^2} \, f(\lambda ), \hskip .2in
f(\lambda) = {2 \over {1- e^{-2ib (\lambda - i \epsilon )}} } 
\label{PHP13}
\eeq
The contour for the $\lambda$-integration is to be completed in the lower half-plane. Using this result in (\ref{PHP10a}), we can carry out the $k_2$-integration. Then we do the $\lambda$-integral to obtain
\begin{align}
{\mathcal M}_\xi&= 2 \pi \, f(- i \omega_p ) \omega_p \delta (p,q)+ 4 p q \int_0^\infty {d s\over \pi} 
\left[ { f(- i \omega_s )\omega_s \over (p^2 -s^2) (q^2 - s^2)} + { f(- i \omega_p )\omega_p \over (s^2 -p^2) (q^2 - p^2)}\right. \nonumber\\
&\hskip 2.7in \left.  + { f(- i \omega_q )\omega_q \over (s^2 -q^2) (p^2 - q^2)}
\right]\label{PHP14a}
\end{align}
where $\omega_p = \sqrt{k_0^2 +k_3^2 +p^2}$, etc., and 
\beq
f(-i \omega ) \omega  = {2 \, \omega \over 1- e^{-2 b \omega}} 
\label{PHP15}
\eeq
The integration over $\phi_1$ produces the additional term
$- N_{1\xi}(-k) ( {\mathcal M}_1)^{-1} N_{1\xi}$, which is written out as
\begin{align}
- N_{1\xi} ( {\mathcal M}_1)^{-1} N_{1\xi} &= -\int {dk_2 \over 2 \pi} 
{4 p q \over [p^2 - (k_2 - i \epsilon )^2 ] [q^2 - (k_2 + i \epsilon )^2]}
{\omega^2 \over (\sinh b\omega )^2 (\omega \coth b \omega + \kappa_I)}
\nonumber\\
&= -\int {d \lambda \over \pi} {\omega^2 \over \lambda^2 + \omega^2} \, h(\lambda)\,
{4 p q \over [p^2 - (k_2 - i \epsilon )^2 ] [q^2 - (k_2 + i \epsilon )^2]}\nonumber\\
h(\lambda)&= {i \lambda \over [ \kappa_I (\sin b(\lambda -i \epsilon))^2 +
(\lambda - i \epsilon ) \sin b (\lambda -i \epsilon) \, \cos b (\lambda - i \epsilon ) ]}
\label{PHP16}
\end{align}
By carrying out the integrations as we did to obtain (\ref{PHP14a}), we find
\begin{align}
- N_{1\xi} ( {\mathcal M}_1)^{-1} N_{1\xi} & = -2 \pi  \, h(-iK_p ) K_p\,\delta (p- q)
- 4 p q \int_0^\infty {d s\over \pi} 
\left[ { h(- i K_s )K_s \over (p^2 -s^2) (q^2 - s^2)} + \right. \nonumber\\
&\hskip .5in \left. + { h(- i K_p )K_p \over (s^2 -p^2) (q^2 - p^2)} + { h(- i K_q )K_q \over (s^2 -q^2) (p^2 - q^2)}
\right]\label{PHP17}
\end{align}
with
\beq
h (-i K) K = {K^2 \over (\sinh bK)^2 ( K \coth bK \, + \kappa_I)}
\label{PHP18}
\eeq
Combining (\ref{PHP14a}) and (\ref{PHP17}) we get
\begin{align}
{\mathbb M}_\xi &= 2 \pi  \, (H_p+\omega_p) \delta (p- q)+ (\Delta {\mathbb M}_\xi)_{pq}\nonumber\\
(\Delta {\mathbb M}_\xi)_{pq}&=  4 p q \int_0^\infty {d s\over \pi} 
\left[ { (H_s+\omega_s) \over (p^2 -s^2) (q^2 - s^2)} + {(H_p+\omega_p)\over (s^2 -p^2) (q^2 - p^2)}
+ { (H_q+\omega_q) \over (s^2 -q^2) (p^2 - q^2)}\right]
\label{PHP25}
\end{align}
In a similar way,
\begin{align}
{\mathbb M}_R&= 2 \pi  \, (\omega_p + \kappa_{R}) \delta (p- q)  + (\Delta {\mathbb M}_R)_{pq}
\nonumber\\
(\Delta {\mathbb M}_R)_{pq}& =  4 p q \int_0^\infty {d s\over \pi} 
\left[ { \omega_s \over (p^2 -s^2) (q^2 - s^2)} + { \omega_p \over (s^2 -p^2) (q^2 - p^2)} + { \omega_q \over (s^2 -q^2) (p^2 - q^2)}
\right]
\label{PHP14b}
\end{align}
We follow a similar strategy for ${\mathcal M}_{2L}$ using the integral representation
\beq
\omega \coth b \omega =
\int {d \lambda \over \pi} {\omega^2 \over \lambda^2 + \omega^2} \,
{\tilde f} (\lambda), \hskip .3in {\tilde f}(\lambda) = -i \cot b (\lambda - i \epsilon)
\label{PHP19}
\eeq
This leads to the expression
\begin{align}
{\mathcal M}_{2L} &= 2 \pi \, (\omega_p\coth b\omega_p + \kappa_{L}) \delta (p- q)
+ 4 p q \int_0^\infty {d s\over \pi} 
\left[ { \omega_s\coth b\omega_s \over (p^2 -s^2) (q^2 - s^2)} + {\omega_p\coth b\omega_p \over (s^2 -p^2) (q^2 - p^2)}\right. \nonumber\\
&\hskip 3in \left.  + { \omega_q\coth b\omega_q  \over (s^2 -q^2) (p^2 - q^2)}\right]
\label{PHP20}
\end{align}
We also get a term $ - N_{1\,2L}(- k) ({\mathcal M}_1)^{-1} N_{1\,2L}$ from
integration over $\phi_1$.
Notice that $N_{1\,2L}$ is the same as $N_{1\xi}$
with $k_2 \rightarrow -k_2$ and an overall minus sign. These do not affect the final expression and we get $ N_{1\,2L}(- k) ({\mathcal M}_1)^{-1} N_{1\,2L} = N_{1\xi} ( {\mathcal M}_1)^{-1} N_{1\xi}$. Combining this with (\ref{PHP20}), we get
\begin{align}
{\mathbb M}_L &=  2\pi\, (H_p+\kappa_L) \, \delta (p- q) + (\Delta {\mathbb M}_L)_{pq}
\nonumber\\
(\Delta {\mathbb M}_L)_{pq}&= 4 pq\,\int_0^\infty {ds\over \pi} 
\left[ {H_s \over (p^2 - s^2) (q^2 - s^2)} + {H_p \over (s^2 - p^2) (q^2 - p^2)}  +{H_q \over (s^2 - q^2) (p^2 - q^2)}  \right]
\label{PHP21x}
\end{align}
The expression for $Q_{2L\xi}$ simplifies as
\beq
Q_{2L\xi}= - 4 p q \int_0^\infty {d s\over \pi} 
\left[ { \omega_s \coth b\omega_s \over (p^2 -s^2) (q^2 - s^2)} + {\omega_p \coth b\omega_p \over (s^2 -p^2) (q^2 - p^2)}
+ { \omega_q\coth b\omega_q \over (s^2 -q^2) (p^2 - q^2)}\right]
\label{PHP21a}
\eeq
which is just the second part of the expression for ${\mathcal M}_{2L}$. 
The integration over $\phi_1$ also produces a new mixing term between $\phi_{2L}$ and
$\xi$ given by $N^T_{1\xi} ({\mathcal M}_1)^{-1} N_{1\,2L}$ which can be simplified as
\begin{align}
N^T_{1\xi} ({\mathcal M}_1)^{-1} N_{1\,2L}&= - 
 4 p q \int_0^\infty {d s\over \pi} 
\left[ { h(- i K_s )K_s \over (p^2 -s^2) (q^2 - s^2)} + { h(- i K_p )K_p \over (s^2 -p^2) (q^2 - p^2)}\right. \nonumber\\
&\hskip 1.2in \left.  + { h(- i K_q )K_q \over (s^2 -q^2) (p^2 - q^2)}
\right] \label{PHP22}
\end{align}
This expression can be combined with (\ref{PHP21a}) to get
\begin{align}
N_{2L\xi} &\equiv (Q_{2L\xi} - N^T_{1\xi} ({\mathcal M}_1)^{-1} N_{1\,2L} )_{pq} \nonumber\\
&=
 - 4 p q \int_0^\infty {d s\over \pi} 
\left[ { H_s  \over (p^2 -s^2) (q^2 - s^2)} + { H_p \over (s^2 -p^2) (q^2 - p^2)}
+ {H_q  \over (s^2 -q^2) (p^2 - q^2)}
\right]\nonumber\\
&= - ( \Delta {\mathbb M}_L)_{pq}
 \label{PHP23}
\end{align}

The expression for $N_{2R\xi}$ from (\ref{PHP9b}) can be simplified as follows,
\begin{align}
N_{2R\xi} &= - 4 p q \int_{-\infty}^\infty {dk_2 \over 2 \pi} {4 p q \over [ (k_2 - i \epsilon )^2 - p^2] \, [
  (k_2 - i \epsilon )^2 - q^2]} ~ \int_{-\infty}^\infty {d\lambda \over \pi} {\omega^2 \over \lambda^2 + \omega^2}
\nonumber\\
&= 4 p q \int {d \lambda \over 2 \pi} {\lambda^2 \over \sqrt{\lambda^2 + \vk^2}\,\,
(\lambda^2 + \omega_p^2 )\, (\lambda^2 + \omega_q^2)}\nonumber\\
&= 4 p q \int {d \lambda \over 2 \pi} {1 \over \sqrt{\lambda^2 + \vk^2}}
\left[ {\omega_p^2\over  (\lambda^2 + \omega_p^2 )} -
{\omega_q^2 \over  (\lambda^2 + \omega_q^2)}\right] {1\over p^2 - q^2}
\nonumber\\
&= 4 p q \left[ {\omega_p^2 \,{\cal I}(\vk, p) - \omega_q^2\, {\cal I}(\vk,q)
\over p^2 - q^2}\right]
\label{PHP51}
\end{align}
where
\beq
{\cal I} (\vk,p) = \int_0^\infty  {d \lambda \over \pi} {1\over \sqrt{\lambda^2 + \vk^2}\,\,
(\lambda^2 + \omega_p^2)}
= {1\over 2\pi} {1\over p \, \omega_p} \log
\left( {\omega_p +p \over \omega_p - p}\right)
\label{PHP52}
\eeq
By rewriting $1/\sqrt{\lambda^2 +\vk^2}$ as we did for the others and simplifying,
we can also see that
\beq
N_{2R\xi} = - (\Delta {\mathbb M}_R)_{pq}
\label{PHP52s}
\eeq

We now turn to the $\rho$-dependent terms.
\vskip .1in\noindent
\underline{$\rho \, \rho$ term}

The kernel ${\mathcal M}_\rho$ is
$\omega \coth b\omega + \omega$; it is
easier to simplify the $\rho\, \rho$-term after the integration over $\phi_1$. This yields
the kernel ${\mathbb M}_\rho = {\mathcal M}_\rho - (\omega\, {\rm csch}\,b\omega )^2 ({\mathcal M}_1)^{-1} = H+\omega$. This gives directly
\beq
{\mathbb M}_\rho =  4 \vert\vk\vert^2 \,\int_{-\infty}^\infty {ds \over 2\pi} \,  {H_s+\omega_s \over \omega_s^4}
~+~ {c_0 \over 2 \vert \vk\vert}
\label{PHP9c}
\eeq
\vskip .1in\noindent
\underline{$\rho \, \xi$ term}

This term has $\omega + \omega \coth b\omega - N_{1\xi} ({\mathcal M}_1)^{-1}
N_{1\rho} = H+\omega$ as the kernel, so
we get
\begin{align}
N_{\rho \xi} &= \int {dk_2 \over 2\pi} {2p \over p^2 - (k_2-i\epsilon )^2} {2 \vert\vk\vert \over
\omega_k^2} \, (H_k+\omega_k )\nonumber\\
&= \int {d\lambda \over \pi} {dk_2 \over 2\pi} 
{4 \vert\vk\vert \,p \over [p^2 - (k_2 - i \epsilon )^2]\, (\lambda^2 +\omega_k^2)} \, \Bigl(f(\lambda ) + h (\lambda )\Bigr)\nonumber\\
&=\int {d\lambda \over 2 \pi} 
{4 \vert\vk\vert \,p \over\sqrt{\lambda^2 +\vk^2}\,\,(\lambda^2 +\omega_p^2)} \, \Bigl(f(\lambda ) + h (\lambda )\Bigr)\nonumber\\
&=\int_0^\infty {ds\over \pi} \int_{-\infty}^\infty {d\lambda \over \pi} 
{4 \vert\vk\vert \,p \over (\lambda^2 +\omega_s^2)\,\,(\lambda^2 +\omega_p^2)} \, \Bigl(f(\lambda ) + h (\lambda )\Bigr)\nonumber\\
&= 4 \vert\vk\vert \,p \int_0^\infty {ds\over \pi} \left[ { H_s +\omega_s \over \omega_s^2}
-  { H_p +\omega_p \over \omega_p^2}\right] {1\over p^2 - s^2}
\label{PHP53} 
\end{align}
\vskip .1in\noindent
\underline{$\phi_{2L}\,\rho$ term}

The kernel for this term is $\omega \coth b\omega + c_{2L} - N_{1\rho}({\mathcal M}_1)^{-1} N_{1\,2L} = H + c_{2L}$ and gives
\begin{align}
N_{2L \rho} &= \int {dk_2 \over 2\pi} {2p \over p^2 - (k_2- i \epsilon )^2} {2 \vert\vk\vert \over
\omega^2} \, (H +c_{2L})  \nonumber\\
&= 4 \vert\vk\vert \, p \left[ \int_0^\infty {ds \over \pi} \left(
{H_s \over \omega_s^2} - {H_p \over \omega_p^2} \right) {1\over p^2 - s^2}
~+~ {c_{2L}\over 2 \vert\vk\vert \, \omega_p^2}\right]
\end{align}
For the first term on the right hand side, we simplified as we did for
$N_{\rho \xi}$ and for the second term, we did the $k_2$ integral completing the contour
in the lower half-plane.
\vskip .1in\noindent
\underline{$\phi_{2R}\,\rho$ term}

This is similar to the $\phi_{2L} \rho$ term, except that we have $\omega + c_{2R}$ instead of
$H + c_{2L}$.
We get
\begin{align}
N_{2R \rho}&= - \int {dk_2 \over 2\pi} {4 \vert \vk\vert \, p\over [p^2 - (k_2 -i\epsilon )^2]
\,\omega^2} \, (\omega + c_{2R})\nonumber\\
&= - \int  {d\lambda \over \pi} {dk_2 \over 2\pi} {4 \vert \vk\vert \, p\over [p^2 - (k_2 -i\epsilon )^2]
\,\omega^2} \, {\omega^2 \over \omega^2 +\lambda^2} - c_{2R}\int {dk_2 \over 2\pi} {4 \vert \vk\vert \, p\over [p^2 - (k_2 -i\epsilon )^2]
\,\omega^2}\nonumber\\
&= - 4 \vert \vk\vert p \int {d\lambda \over 2\pi} {1\over \sqrt{\lambda^2 + \vk^2}\,\, (\lambda^2 +
\omega_p^2)} - c_{2R} {2 p \over \omega_p^2}
\nonumber\\
&= - 4 \vert\vk\vert \, p \left[ {\cal I}(\vk,p) 
~+~ {c_{2R}\over 2 \vert\vk\vert \, \omega_p^2}\right]
\end{align}

Finally, we note that in the calculation of the Casimir energy by expansion of the 
$\Tr \log$, the first order correction involves the diagonal elements of the type
$\Delta_{pp}$. In some cases (such as $\Delta {\mathbb M}_L$, $\Delta {\mathbb M}_\xi$, etc.)
$\Delta_{pq}$ is of the form
\beq
\Delta_{pq} =  4 pq\,\int_0^\infty {ds\over \pi} 
\left[ {W_s \over (p^2 - s^2) (q^2 - s^2)} + {W_p \over (s^2 - p^2) (q^2 - p^2)}  +{W_q \over (s^2 - q^2) (p^2 - q^2)}  \right]
\eeq
The $p =q$ limit can be extracted as
\beq
\Delta_{pp} = 4\, p^2 \int {ds \over \pi} \left[
{{W_s - W_p} \over (s^2 - p^2)^2} -  \left( { \del W \over \del p^2}\right) {1\over (s^2 - p^2)}
\right]
\label{PHP37}
\eeq
We will need this expression with the appropriate $W$'s to estimate the contribution to the energy numerically.


\end{document}